\documentclass[11pt,a4paper]{article}
\pdfoutput=1

\usepackage{jheppub}
\usepackage{color,graphicx} \usepackage{ifpdf}
\usepackage{amsmath} \usepackage{amssymb} \usepackage{bm}

\usepackage[utf8]{inputenc} 
\usepackage[T1]{fontenc}

\usepackage{hhline}
\usepackage{color}
\definecolor{nicered}{rgb}{0.7,0.1,0.1}
\definecolor{nicegreen}{rgb}{0.1,0.5,0.1}

\usepackage[english]{babel}
\usepackage{slashed}
\usepackage{multirow}
\usepackage{braket}
\usepackage{slashed}

\newcommand {\E}[1]{\times 10^{#1}}	
\newcommand {\e}[1]{\mathrm{\,#1}}       
\newcommand{\mc}[1]{\mathcal{#1}}

\newcommand{\mrm}[1]{\mathrm{#1}}
\newcommand{\re}[0]{\mrm{Re}}

\newcommand{\bea}{\begin{eqnarray}}
\newcommand{\eea}{\end{eqnarray}}

\newcommand{\br}[0]{\mc{B}}

\definecolor{Red}{rgb}{1.,0.,0.}

\definecolor{Green}{rgb}{0.2,.7,0.2}

\usepackage{mciteplus}

\usepackage{hyperref}
\hypersetup{colorlinks,citecolor=nicegreen,linkcolor=nicered}
\hypersetup{colorlinks=true}

\begin{document}





\arxivnumber{1711.07779}

\author[a,c]{Ilja Dor\v sner} 
\author[b,c]{Svjetlana Fajfer} 
\author[c]{Darius A. Faroughy}
\author[b,c]{Nejc Ko\v snik} 

\emailAdd{ilja.dorsner@ijs.si}
\emailAdd{svjetlana.fajfer@ijs.si}
\emailAdd{nejc.kosnik@ijs.si}
\emailAdd{darius.faroughy@ijs.si}

\affiliation[a]{University of Split, Faculty of Electrical Engineering, Mechanical Engineering and Naval Architecture in Split (FESB), Ru\dj era Bo\v skovi\' ca 32, 21000 Split, Croatia}
\affiliation[b]{Department of Physics,
  University of Ljubljana, Jadranska 19, 1000 Ljubljana, Slovenia}
\affiliation[c]{J. Stefan Institute, Jamova 39, P. O. Box 3000, 1001
  Ljubljana, Slovenia}

\title{The role of the $S_3$ GUT leptoquark in flavor universality and
  collider searches}

\date{\today}

\abstract{ We investigate the ability of the $S_3$ scalar
  leptoquark to address the recent hints of lepton universality
  violation in $B$ meson decays. The $S_3$ leptoquark with
  quantum numbers $(\overline{\bm{3}},\bm{3},1/3)$
  naturally emerges in the context of an $SU(5)$ GUT model
  without any conflict with the stringent limits from observed nucleon
  stability. Scalar leptoquark $S_3$ with left-handed couplings to 2nd
  and 3rd generations of charged leptons and down-type quarks seems
  well-suited to address both $R_{K^{(*)}}$ and $R_{D^{(*)}}$. We
  quantify this suitability with numerical fits to a plethora of
  relevant flavor observables. The proposed
  $SU(5)$ model calls for a second leptoquark state, i.e., $\tilde{R}_2$ with quantum numbers
  $(\bm{3},\bm{2},1/6)$, if one is to generate gauge coupling
  unification and neutrino mass. We accordingly include it in our study to investigate
  $\tilde{R}_2$'s ability to offset adverse effects of $S_3$ and thus
  improve a quality of numerical fits. A global fit of the leptoquark
  Yukawa couplings shows that large couplings of light $S_3$ to $\tau$
  leptons are preferred.
We furthermore identify $B \to K^{(*)} \bar\nu\nu$ as the most
  sensitive channel to probe the preferred region of parameter
  space. Large couplings of $S_3$ to $\tau$ leptons are finally
  confronted with the experimental searches for $\tau$ final states at
  the Large Hadron Collider. These searches comprise a study of decay
  products of the leptoquark pair production, as well as, and more
  importantly, an analysis of the high-mass $\tau\tau$ final states.
}

\maketitle

 \section{Introduction}

 At low energies there are a few experimentally measured observables that exhibit deviation from
 the Standard Model (SM) predictions. Among them, the three $B$ meson
 anomalies, indicating possible lepton flavor universality~(LFU)
 violation, particularly stand out.
 
One of these anomalies manifests itself in the ratios
\begin{align}
R_{D^{(*)}} = \frac{\Gamma(B \to D^{(*)} \tau^- \bar\nu)}{\Gamma(B \to  D^{(*)} \ell^- \bar\nu)}\, ,
\end{align}
according to the experimental results in
Refs.~\cite{Lees:2012xj,Lees:2013uzd, Huschle:2015rga, Adachi:2009qg,
  Bozek:2010xy, Aaij:2015yra,Hirose:2016wfn}. The result for $R_D $
appears to be $1.9\,\sigma$ larger than the SM prediction,
i.e., $R_D^{\rm SM}=0.286\pm 0.012$, that is obtained by relying on the lattice QCD
results for both the vector and the scalar form factors
\cite{Becirevic:2016yqi} (see also Ref.~\cite{Bigi:2016mdz}). The
experimentally established $R_{D^\ast}=0.304\pm 0.020$ has also been
confirmed~\cite{Aaij:2015yra,Hirose:2016wfn,Amhis:2016xyh}, and it appears to be
$\sim 3\,\sigma$ larger than predicted value
$R_{D^\ast}^{\rm SM}=0.252\pm 0.003$~\cite{Fajfer:2012jt}.  The
deviation from the SM prediction in the $R_D$--$R_{D^*}$ plane is at
3.9\,$\sigma$ level~\cite{Ligeti:2016npd,Crivellin:2016ejn, Amhis:2016xyh} and it has
accordingly attracted a lot of attention
recently~\cite{Altmannshofer:2017poe,Alonso:2016oyd,Crivellin:2014zpa,Bhattacharya:2014wla,Bhattacharya:2015ida,Hati:2015awg,Sakaki:2014sea,Fajfer:2012jt}.

The remaining two $B$ meson anomalies are related to the  $b \to s \ell^+ \ell^-$ transition. Namely, the
LHCb experiment has found that there are slight discrepancies between the SM prediction and the experimental results for the angular
observable known as $P_5^{\prime}$ in $B \to K^\ast \mu^+ \mu^-$ process. In many approaches this disagreement has
been attributed to new physics (NP), although the tension might be a result of
the SM QCD effects (see e.g.\ Ref.~\cite{Capdevila:2017ert} and references therein). The second of the two $b \to s \ell^+ \ell^-$
transition anomalies has been found in the ratio of the branching fractions,
\begin{equation}
\label{e1}
\begin{split}
R_K \equiv \frac{\br( B^+ \to K^+ \mu^+ \mu^-)_{q^2 \in
    [1,6]\e{GeV^2}}}{\br( B^+ \to K^+ e^+ e^-)_{q^2 \in
    [1,6]\e{GeV^2}}} &=  0.745 \pm^{0.090}_{0.074} \pm 0.036~\textrm{\cite{Aaij:2014ora}},\\ 
R_{K^*}^{q^2 \in [1.1,6]\e{GeV}^2} \equiv \frac{\br( B^0 \to K^{*0} \mu^+ \mu^-)_{q^2 \in
    [1.1,6]\e{GeV^2}}}{\br( B^0 \to K^{*0} e^+ e^-)_{q^2 \in
    [1.1,6]\e{GeV^2}}} &=
  0.69^{+0.11}_{-0.07}\pm 0.05~\textrm{\cite{Aaij:2017vbb}},\\
R_{K^*}^{q^2 \in [0.045,1.1]\e{GeV^2}} &=
  0.66^{+0.11}_{-0.07}\pm 0.03~\textrm{\cite{Aaij:2017vbb}}.
\end{split}
\end{equation}
The values the LHCb experiment measured for these ratios are
consistently lower than the SM prediction,
i.e., $R_K^\mrm{SM} =1.00 \pm 0.03$, in which the next-to-next-to-leading QCD
corrections and soft QED effects have been
included~\cite{Bordone:2016gaq,Hiller:2003js} (for $R_{K^*}$ see Table~1 and
references in~\cite{Aaij:2017vbb}). In other words, the
LHCb results point towards a significant effect of the lepton
flavor universality violation in this process. Recently, Belle
Collaboration~\cite{Wehle:2016yoi} found out that the angular
observable $P_5^{\prime}$ agrees with the SM prediction much better for
electrons than for muons. This important result suggests that it is
much more likely that beyond the SM effects are present in the second
generation of leptons, and that there are currently no effects in $b
\to s e^+ e^-$ which would not be accounted for in the SM.

Many scenarios of NP~\cite{Altmannshofer:2014cfa,
  Becirevic:2016yqi,Datta:2013kja,
  Hiller:2014yaa,Crivellin:2014zpa,Glashow:2014iga,Bhattacharya:2014wla,Gripaios:2014tna,Greljo:2015mma,Ghosh:2014awa,
  Crivellin:2015mga,Crivellin:2015lwa,Crivellin:2017zlb,
  Sierra:2015fma,Varzielas:2015iva, Crivellin:2015era,
  Celis:2015ara,Freytsis:2015qca,Fajfer:2015ycq,Cox:2016epl,Becirevic:2017jtw,Kamenik:2017tnu,Arnan:2017lxi,Ghosh:2017ber,Bardhan:2016uhr,Capdevila:2017bsm}
have been investigated in order to explain either $R_{K^{(*)}}$ and
$P_5^\prime$, or $R_{D^{(*)}}$ anomalies.  An interesting observation
was found in Ref.~\cite{Bhattacharya:2014wla} that $R_{K^{(*)}}$ and
$P_5^\prime$ can be explained if NP couples only to the third
generation of quarks and leptons.  Furthermore, the authors of
Refs.~\cite{Calibbi:2015kma,Greljo:2015mma} noticed that both
$R_{D^{(*)}}$ and $R_{K^{(*)}}$ puzzles can be correlated if the
effective four-fermion semileptonic operators consist of left-handed
doublets.

In this work we consider a Grand Unified Theory~(GUT) inspired setting
with a light scalar $S_3$ leptoquark (LQ) 
that transforms as $(\overline{\bm{3}},\bm{3},1/3)$ 
under the SM gauge group $SU(3) \times SU(2) \times U(1)$. The state $S_3$ is rendered baryon
number conserving due to the GUT symmetry, as we discuss in
Sec.~\ref{sec:gut}, and generates purely left-handed current
$\bar L L \bar Q Q$ operators which seem to be well-suited to explain the LFU puzzles
if the $S_3$ mass is at the TeV scale. The need for the second light
LQ state, $\tilde R_2$ in representation $(\bm{3},\bm{2},1/6)$,
emerges naturally from the requirement of neutrino masses generation
in the advocated GUT model as well as from the gauge coupling unification. We accordingly include  $\tilde R_2$ in our study and investigate whether it could partially compensate for the adverse low-energy effects of $S_3$. In Sec.~\ref{sec:setup} we
introduce relevant couplings of these two LQs with the SM fermions. We proceed to show how $S_3$ could, in principle, address the LFU puzzles
in Sec.~\ref{sec:lfuv}. We then present relevant additional constraints on
the LQ parameters in Sec.~\ref{sec:LQcons}. The low-energy
flavor analysis is concluded in
Sec.~\ref{sec:fit} with the global fit of the relevant couplings of the two LQs with quark-lepton pairs for three specific Yukawa structures. Sec.~\ref{sec:collider} is devoted to collider
study of the model signatures in the LQ resonant pair production and
in a $t$-channel LQ exchange contributing to $\tau\tau$ final states
at LHC. We elaborate on the GUT construction behind the two LQ states in
Sec.~\ref{sec:gut}. Finally, we conclude in Sec.~\ref{sec:conclusion}.

\section{Model setup}
\label{sec:setup}

The LQ multiplet $S_3 (\overline{\bm{3}},\bm{3},1/3)$
interacts with the SM fermions in accordance with
its quantum numbers, given in the brackets. The three charge eigenstate 
components of $S_3$, i.e., $S^{4/3}_3$, $S^{1/3}_3$, and $S^{-2/3}_3$, have
the following Yukawa interactions with fermions~\cite{Dorsner:2016wpm}
\begin{equation}
  \label{eq:S3}
\begin{split}
\mathcal{L}_{S_3} = &-y_{ij}\bar{d}_{L}^{C\,i} \nu_{L}^{j}
S^{1/3}_{3}-\sqrt{2} y_{ij}\bar{d}_{L}^{C\,i} e_{L}^{j} S^{4/3}_{3}+\\
&+\sqrt{2} (V^* y)_{ij}\bar{u}_{L}^{C\,i} 
    \nu_{L}^{j} S^{-2/3}_{3}-(V^* y)_{ij}\bar{u}_{L}^{C\,i} 
    e_{L}^{j} S^{1/3}_{3} + \mrm{h.c.},
\end{split}
\end{equation}
where $V$ is the Cabibbo-Kobayashi-Maskawa (CKM) mixing matrix. Note
that $S_3$ has purely left-handed couplings. The diquark interactions
with $S_3$ are not shown in Eq.~\eqref{eq:S3} since we assume that
$S_3$ and its interactions originate from the GUT construction
presented in Ref.~\cite{Dorsner:2017wwn} where the baryon number
violating diquark couplings are forbidden due to the grand unified
symmetry.\footnote{Complete model-independent sets of $S_3$ and
  $\tilde R_2$ couplings to fermions can be found
  in Ref.~\cite{Dorsner:2016wpm}.} The main goal of our study is to address
the puzzles observed in neutral current LFU tests in the $R_K$ ratio
(and related anomalies in $b\to s \mu^+ \mu^-$) as well as in
charged-current LFU ratios $R_{D^{(*)}}$. Thus we have clear target
observables that we can affect with a small number of LQ Yukawa
couplings. 

In the context of SM complemented with effective operators
(SM-EFT) it has been shown that NP models contributing to dimension-6
operators made out of left-handed quark and lepton doublets can
explain both neutral- and charged-current LFU
anomalies~\cite{Bhattacharya:2014wla,Alonso:2015sja, Calibbi:2015kma,Greljo:2015mma,Feruglio:2017rjo}.
However, in an explicit NP model these effective interactions could be correlated, unlike in
the effective theory\footnote{Even in the effective theory the
quantum corrections have strong effect on low-energy precision
measurements~\cite{Feruglio:2016gvd,Feruglio:2017rjo}.},
with other observables. Our intention is to quantify this correlation within this particular model.

The LQ state $S_3$ can affect all the target LFU observables
with a minimal set of parameters, e.g., $y_{s\mu}$, $y_{b \mu}$, and
$y_{b \tau}$. In this work, however, we also study the effect of the coupling
$y_{s\tau}$ which enables a handle on the semitauonic modes entering
$R_{D^{(*)}}$. The couplings of $S_3$ to $d_L$ and $e_L$ have to be rather small in order to avoid pressing bounds from LFV and kaon physics. We opt to set those couplings to zero to obtain the following flavor structure:
\begin{equation}
\label{eq:textureS3}
 y =
\begin{pmatrix}
  0 & 0 & 0\\
  0 & y_{s \mu} & y_{s \tau}\\
  0 & y_{b \mu} & y_{b \tau}
\end{pmatrix},
\qquad
 {V^*} y=
\begin{pmatrix}
  0 &V_{us}^* y_{s\mu} +V_{ub}^* y_{b\mu} &V_{us} ^*y_{s\tau} +V_{ub} ^*y_{b\tau}\\
  0 &V_{cs}^* y_{s\mu} +V_{cb}^*y_{b\mu} &V_{cs} ^*y_{s\tau} +V_{cb} ^*y_{b\tau}\\
  0 &V_{ts}^* y_{s\mu} +V_{tb} ^*y_{b\mu} &V_{ts}^* y_{s\tau}+V_{tb}^* y_{b\tau}
\end{pmatrix}.
\end{equation}
Note that the Yukawa couplings of $S_3$ to up-type quarks are
spread over generations due to CKM rotation. In what follows all Yukawa couplings are assumed to be real. The ansatz of Eq.~\eqref{eq:textureS3} summarizes the most general $S_3$ scenario studied within our work, although we will also comment on more restricted scenarios, where some additional elements of $y$ will be set to zero.  

Having only one LQ with
mass around the $1\e{TeV}$ scale would invalidate unification of gauge
couplings, thus a second LQ state --- $\tilde R_2$ in
our case --- is needed. The two electric charge eigenstates of
$\tilde R_2$ couple only to down-type quarks:
\begin{equation}
  \label{eq:R2}
\begin{split}
  \mathcal{L}_{\tilde R_2} =&  -\tilde{y}_{ij}\bar{d}_{R}^{i}e_{L}^{j}\tilde{R}_{2}^{2/3}+\tilde{y}_{ij}\bar{d}_{R}^{i}\nu_{L}^{j}\tilde{R}_{2}^{-1/3}+\textrm{h.c.}.
\end{split}  
\end{equation}
The doublet $\tilde R_2$ can accomodate the measured value of $R_K$, but its
right-handed current contributions cause tensions with the reported
value for $R_{K^*}$. In the current setting with strictly left-handed
neutrinos $\tilde R_2$ does not interact with up-type quarks and
thus cannot affect $R_{D^{(*)}}$. In our approach it is $S_3$ that
could, in principle, address both LFU anomalies, whereas its side-effects in other
well-constrained observables (e.g. $B_s$--$\bar B_s$ mixing and $B \to
K^{(*)} \bar \nu \nu$) might be, hopefully, cancelled by $\tilde R_2$. 
Since $S_3$ will have largest effects
in the $\tau$ sector we have to introduce couplings of $\tilde R_2$
to $\tau$ in order to compensate for potentially unwanted effects. In the following
analysis we will analyze a light $S_3$ scenario with the couplings
texture~\eqref{eq:textureS3} and along with it test the viability of
having light $\tilde R_2$ with nonzero Yukawas involving the $\tau$ lepton.
Namely, we take
\begin{equation}
  \label{eq:textureR2}
\tilde y =
\begin{pmatrix}
  0 & 0 & 0\\
  0 & 0 & \tilde y_{s \tau}\\
  0 & 0 & \tilde y_{b \tau}
\end{pmatrix}.
\end{equation}
The mass of $\tilde R_2$ should be at around $1$\,TeV in order to affect low-energy phenomenology, if required at all. We consistently take this to be the case when we discuss the role of
$\tilde R_2$ in gauge coupling unification and the neutrino mass generation.

For both LQ states the rotations with the CKM matrix $V$, left over from the transition to the
mass basis of fermions, have been assigned to the $u_L$ fields. For the
study of flavor phenomenology the neutrinos can be safely considered
as massless. Thus, Lagrangians in Eqs.~\eqref{eq:S3} and \eqref{eq:R2} are
written in the fermion mass basis with the exception of $\nu_L$ whose
mass basis is ill-defined. We use flavor basis for the neutrinos, such
that the Pontecorvo-Maki-Nakagawa-Sakata (PMNS) matrix becomes unity.

\section{LFU violating contributions}
\label{sec:lfuv}

In this section we focus on how the two light LQs would affect
the LFU violating anomalies measured in $B$ meson decays. The gross features
required of Yukawa matrices will be presented. The detailed discussion of additional observables and their interplay with the LFU anomalies will be
presented in the next section.

\subsection{Charged currents LFU: $R_{D^{(*)}}$}

The largest LFU violating effect is in the charged current observables
$R_{D^{(*)}}$. For a NP-induced effective operator that follows
the chirality structure of the SM it has been shown that the
dimensionless coupling of $\sim 0.1$ is needed, if new particles have
mass of $\Lambda = 1\e{TeV}$ and contribute at
tree level~\cite{Freytsis:2015qca}. The matched contributions of
$S_3$ generate left-handed current operator, whereas $\tilde R_2$
cannot contribute to charged currents in this setup\footnote{Charged
  currents can be induced by $\tilde R_2$ if right-handed neutrinos
  are added to the fermion sector.}. In particular in
$b \to c \ell \bar\nu$ transition the $S_3$ presence leads to the
modification of the left-handed charged-current operators:
\begin{equation}
  \label{eq:ccLag}
\mc{L}_\mathrm{SL} = -\frac{4G_F}{\sqrt{2}}
  \Bigg[ (V_{UD} +g^{L}_{UD;\ell \ell})(\bar U
  \gamma^\mu P_L D) (\bar \ell \gamma_\mu P_L
  \nu_{\ell}) \Bigg], \qquad U=u,c,t,\quad D=s,b,\quad \ell=\mu, \tau,
  \end{equation}
where the LQ  term in
Eq.~\eqref{eq:ccLag} reads
\begin{equation}
\label{eq:gUDlnu}
g_{UD;\ell \nu}^L = -\frac{v^2}{4 m_{S_3}^2} (V y^* )_{U \ell}
y_{D\nu}.
\end{equation}
The effect of $S_3$ may be also understood as
 nonuniversal CKM elements in semileptonic charged-current processes:
\begin{equation}
  \label{eq:CKMresc}
  |V_{ij}^{(\ell)}|^2 = |V_{ij}|^2 \left[1 - \frac{v^2}{2m_{S_3}^2} \re
    \left(\frac{V_{is}}{V_{ij}} y_{s\ell}^* y_{j \ell} +
      \frac{V_{ib}}{V_{ij}} y_{b\ell}^* y_{j\ell}\right)\right],\qquad
  i = u,c,t,\quad j=s,b,\quad \ell = \mu,\tau.
\end{equation}
One also has lepton flavor violating $S_3$ contributions
parameterized by $g_{UD;\ell \nu}$, with their effect being much smaller
since they do not interfere with the SM amplitude. They contribute at
subleading order, namely at $v^4/m_{S_3}^4$ that we neglect in
comparison to the interference terms. Here $v=246\e{GeV}$ is the electroweak vacuum expectation value. Notice that the
form of interaction imposed in Eq.~\eqref{eq:textureS3} implies that both
decay modes $B \to D^{(*)} \tau \nu_\tau$ and
$B \to D^{(*)} \mu \nu_\mu$ are affected. From the fit to the measured
ratio $R_{D^{(*)}}$, performed in Ref.~\cite{Freytsis:2015qca}, we
learn that at $1\,\sigma$ we have the following constraint on the
$S_3$ Yukawas:
\begin{equation}
\label{eq:RDcons}
\re\left[V_{cb} \left(|y_{b\tau}|^2 - |y_{b\mu}|^2 \right)  + V_{cs}
  \left(   y_{b\tau} y_{s\tau}^* 
-
  y_{b\mu} y_{s\mu}^* \right) \right]= -2 C_{V_L}
\left(m_{S_3}/\mrm{TeV}  \right)^2,\quad C_{V_L} = 0.18 \pm
0.04\,.
\end{equation}
The $R_{D^{(*)}}$ constraint of Eq.~\eqref{eq:RDcons} includes effects
from $\tau \bar \nu_\tau$ and $\mu \bar \nu_\mu$ states. It is
important to notice definite signs of the LQ-SM interference
contributions which are proportional to
$V_{cb}$. Large $y_{b\tau}$ is clearly disfavoured
by~\eqref{eq:RDcons} while $y_{b\mu}$ results in negative interference
term in semi-muonic modes that would be welcome from the $R_{D^{(*)}}$
point of view, however this possibility could
be in conflict with precise measurements of LFU in
$R_{D^{(*)}}^{\mu/e}$ (studied below in Sec.~\ref{sec:LQcons}). Out of the remaining
two terms $y_{b\mu} y_{s\mu}^*$ is negligible in Eq.\eqref{eq:RDcons} as required by the
$b\to s\mu^+ \mu^-$. The only numerical scenario with positive
interference term for the semi-tauonic mode is the one with large Cabibbo favored
contribution,
\begin{equation}
  \label{eq:RDconsSimple}
y_{b\tau} y_{s\tau}^* \approx -0.4 (m_{S_3}/\mrm{TeV})^2.
\end{equation}
In the next section we will introduce constraints that put important
bound on the above product of Yukawas.

\subsection{Neutral currents: $R_{K^{(*)}}$, $\br(B \to K^{(*)} \mu^+ \mu^-)$ and
  related observables}
The $R_K$ anomaly can be accounted for by the additional contribution
of $S_3$ state to the effective four-Fermi operators that are a
product of left-handed quark and lepton
currents~\cite{Dorsner:2016wpm}.  The $\tilde R_2$ state alone can
also explain $R_K$ via the right-handed current
operators~\cite{Becirevic:2015asa}, but the recent measurement of
$R_{K^*}$ being significantly smaller than 1~\cite{Aaij:2017vbb}
implies that these operators' contributions must be
small~\cite{Hiller:2014yaa, Becirevic:2015asa}. If we expand our
analysis to a whole family of observables driven by
$b \to s \mu^+ \mu^-$ process the scenario with left-handed currents
($S_3$ state) presents a good fit and prefers the following range
at $1\,\sigma$~\cite{Capdevila:2017bsm} (see
also~\cite{Alonso:2015sja,Descotes-Genon:2015uva}):
\begin{equation}
  \label{eq:C9fit}
C_9 = -C_{10} = -0.61^{+0.13}_{-0.10}.
\end{equation}
The exchange of $S_3^{4/3}$ contributes towards the above effective
coefficients as
\begin{equation}
\label{eq:C9}
C_9 = -C_{10} =  \frac{\pi}{V_{tb}V_{ts}^* \alpha}\, y_{b \mu} y_{s \mu}^* \frac{v^2}{m_{S_3}^2}.
\end{equation}
For a
range~\eqref{eq:C9fit} of Wilson coefficients we find
\begin{equation}
\label{eq:bsmumuCons}
y_{b \mu} y_{s \mu}^* = (-0.958^{+0.016}_{-0.020})  \E{-3} \,\left(m_{S_3}/{\rm TeV}
\right)^2.
\end{equation}
Contrary to $S_3$, the right-handed quark currents generated by
$\tilde R_2$ do not improve significantly the global agreement
between theory predictions and observables related to the
$b \to s \mu^+\mu^-$. Tree-level matching of $\tilde R_2$ amplitudes
yields
\begin{equation}
  \label{eq:C9prime}
  C_9' = -C_{10}' = -\frac{\pi}{V_{tb} V_{ts}^* \alpha}\, \tilde y_{s
    \mu} \tilde y_{b \mu}^* \frac{v^2}{m_{\tilde R_2}^2}.
\end{equation}
Using the result of the global fit from~\cite{Capdevila:2017bsm} we have checked that including non-zero $\tilde y_{s\mu}$ and $\tilde
y_{b\mu}$ does not improve the fit considerably.

\section{Constraints on the LQ couplings}
\label{sec:LQcons}

The introduction of the two LQ states with sizable couplings to
explain the LFU observables, as presented above, inevitably causes
side effects in related flavor observables which we will focus on in
this section.

\subsection{LFU ratios and decay rates in charged currents}
\subsubsection{Semileptonic $B$ decays}
Besides measuring $R_{D^{(*)}}$ that does not distinguish between $e$ and
$\mu$ in the final state, Belle Collaboration also reported on the
lepton universality ratio in $e$ and $\mu$. Here we will use $R_{D^*}^{e/\mu} =
1.04(5)(1)$~\cite{Abdesselam:2017kjf} and $R_{D}^{\mu/e} =
0.995(22)(39)$~\cite{Glattauer:2015teq}, both of which are consistent
with $1$. In our framework the $S_3$ state can potentially contribute
to those ratios by rescaling the overall normalization of $B \to
D^{(*)} \mu \nu$. It follows from Eq.~\eqref{eq:CKMresc} that the $S_3$
contributions in these decays are constrained:
\begin{equation}
  \label{eq:RDmue}
  -\frac{v^2}{2m_{S_3}^2} \re\left[\left(\frac{V_{cs}}{V_{cb}}
      y^*_{s\mu} + y^*_{b \mu}\right) y_{b \mu} \right] = R_{D^{(*)}}^{\mu/e} - 1
  = -0.023\pm 0.043,
\end{equation}
where we have averaged over the two Belle results. Due to its
smallness the term $y_{s\mu} y^*_{b\mu}$ is irrelevant in the above
equation (see Eq.~\eqref{eq:bsmumuCons}), albeit the factor $\sim 20$
enhancement due to CKM. After this simplification Eq.~\eqref{eq:RDmue} becomes a rather weak limit, i.e.,
$|y_{b\mu}| \lesssim 1.5 (m_{S_3}/\mrm{TeV})$. It is, however, clear
that $y_{b\mu}$, in spite of its large value, is not sufficient to
explain the $R_{D^{(*)}}$ constraint of Eq.~\eqref{eq:RDcons}. Since the
largest effects are concentrated in the $\tau$ flavor, we expect large
effect in leptonic decay of $B^- \to \tau \bar \nu$ which is sensitive
to
$|V_{ub}^{(\tau)}|^2 \approx |V_{ub}|^2 [1-v^2/(2m_{S_3}^2)
\re((V_{us}/V_{ub}) y_{s\tau}^* y_{b\tau})]$, as given in
Eq.~\eqref{eq:CKMresc}. The $B^-\to \tau \bar \nu$ rate is thus
enhanced by the same combination of Yukawas (and same order of Cabibbo
angle) that also drives the $B\to D^{(*)} \tau \bar \nu$ rate. The
current experimental average
$\br(B^- \to \tau \bar \nu) = (1.09\pm 0.24)\E{-4}$ is indeed slightly
higher than the SM prediction
$\br(B^- \to \tau \bar \nu)^\mrm{SM} = (0.78\pm 0.07)\E{-4}$. 
If we assume that LQ Yukawas are real numbers then the leading contribution
$y_{s\tau}^* y_{b\tau}$ in both observables leads to correlation
\begin{equation}
\label{eq:Btaunu}
  \frac{\br(B^- \to \tau \bar \nu)}{\br(B^- \to \tau \bar
    \nu)^\mrm{SM}} - 1 \approx \left(\frac{R_{D^{(*)}}}{R_{D^{(*)}}^\mrm{SM}} - 1\right) \frac{\rho}{\rho^2+\eta^2},
\end{equation}
where the CKM factor relating the two observables is
close to unity.

\subsubsection{Semileptonic $K$ and $\tau$ decays}
On the other hand, LFU in kaon decays has been tested and confirmed with
 better precision through the following ratios:
\begin{equation}
R_{e/\mu}^K  = \frac{\Gamma(K^- \to e^- \bar \nu)}{\Gamma(K^- \to \mu^-
  \bar \nu)}, \qquad
R_{\tau/\mu}^K  = \frac{\Gamma(\tau^-\to K^- \nu)}{\Gamma(K^- \to \mu^- \bar \nu)}.
\end{equation}
As pointed out in Ref.~\cite{Fajfer:2015ycq} these observables enable us to put strong constraints on the corrections arising within models of NP.
In the $e/\mu$ sector the experimental result~\cite{Olive:2016xmw} agrees well with the SM prediction~\cite{Cirigliano:2007xi}:
\begin{equation}
\label{eq:Remu}
  R_\mrm{e/\mu}^{K(\mrm{exp})} = (2.488\pm 0.010)\E{-5},\qquad
  R_\mrm{e/\mu}^{K(\mrm{SM})} =(2.477\pm 0.001)\E{-5}.
\end{equation}
Using Eq.~\eqref{eq:CKMresc} we recast Eq.~\eqref{eq:Remu}:
\begin{equation}
  \begin{split}
  \label{eq:RKemu}
  \frac{R_\mrm{e/\mu}^{K(\mrm{exp})}}{R_\mrm{e/\mu}^{K(\mrm{SM})}} -1=
\frac{v^2}{2 m_{S_3}^2}
  \re\left[|y_{s\mu}|^2  + (V_{ub}/V_{us}) y_{b\mu}^* y_{s\mu}\right]
  &= (4.4\pm 4.0)\E{-3} \\ 
\Rightarrow \quad |y_{s\mu}| &\lesssim   0.5 (m_{S_3}/\mrm{TeV}).
  \end{split}
  \end{equation} 
$R_\mrm{e/\mu}^{K}$ is most sensitive to  $|y_{s\mu}|$ since the
product $y_{b\mu}^* y_{s\mu}$ must be small as dictated by $b\to
s\mu\mu$ sector and comes with an additional CKM suppression.
The agreement of experiment~\cite{Olive:2016xmw} with the SM
prediction~\cite{Pich:2013lsa} in the $\tau/\mu$ exhibits a $\sim 2\,\sigma$ tension:
\begin{equation}
  R_\mrm{\tau/\mu}^{K(\mrm{exp})} = 467.0 \pm 6.7,\qquad
  R_\mrm{\tau/\mu}^{K(\mrm{SM})} =  \frac{m_K^3
    (m_\tau^2-m_K^2)^2}{2 m_\tau m_\mu^2 (m_K^2-m_\mu^2)^2}(1+\delta
  R_{\tau/K}) = 480.3 \pm 1.0,
\end{equation}
where the dominant error of the experimental ratio is due to
the $\tau$ lifetime uncertainty, whereas on the theory side it is the
radiative correction $\delta R_{\tau/K} = (0.90 \pm
0.22)\%$~\cite{Decker:1994ea} which is the source of uncertainty. The constraint is expressed as:
\begin{equation}
\label{eq:Rtaumu}
  \frac{R_\mrm{\tau/\mu}^{K(\mrm{exp})}}{R_\mrm{\tau/\mu}^{K(\mrm{SM})}}-1 =
\frac{v^2}{2 m_{S_3}^2} 
  \re\left[|y_{s\mu}|^2  - |y_{s\tau}|^2  + (V_{ub}/V_{us})
    (y_{b\mu}^* y_{s\mu} -y_{b\tau}^* y_{s\tau}) \right] = (-2.8 \pm 1.4)\E{-2}.
\end{equation}

\subsubsection{Leptonic decays: $W \to \tau \bar \nu$, $\tau \to \ell \bar \nu \nu$}
The SM tree-level vertex $\bar \tau \nu W$ is rescaled due to
penguin-like contribution of both $S_3$ and $\tilde R_2$. As we
integrate out $S_3$ and $\tilde R_2$ at the weak scale the $W$ vertex with $\tau$
leptons reads $\tfrac{-g}{\sqrt{2}} \bar \nu_\tau \slashed{W} P_L \tau
  (1+ \delta_W^{(\tau)})$, where 
\begin{equation}
  \label{eq:Wtaunu}
  \begin{split}
\delta_W^{(\tau)} &= \frac{N_c}{288\pi^2} \left[(2x+6x\log x-6x\pi
      i) \,(|y_{b\tau}|^2+|y_{s\tau}|^2) +\tilde x\,(|\tilde
    y_{s\tau}|^2 + |\tilde
    y_{b\tau}|^2)\right], \\
  x&=\frac{m_W^2}{m_{S_3}^2}, \qquad \tilde x = \frac{m_W^2}{m_{\tilde R_2}^2}.
  \end{split}
\end{equation}
Free color index in the loops graphs results in the $N_c=3$ factor in
front. We have neglected the quark masses in the above calculation and
presented only the leading terms in $x$ and $\tilde x$. The contribution of $S_3$
with mass of $1\e{TeV}$ shifts the $W \to \tau \nu$ decay width 
relatively by $4\E{-4} (|y_{b\tau}|^2+|y_{s\tau}|^2)$ which is well
below the current $\sim 2\%$ experimental precision. The
$W \to \mu \bar \nu$ is also rescaled by an analogous
$\delta_W^{(\mu)}$ factor.

At low energies the effective $W \to \tau \nu$ vertex would, together
with direct box contributions with LQs, manifest itself in the
$\tau \to \ell \bar \nu_\ell \bar \nu_\tau$ decays. Only $S_3$ may
participate in the box diagrams since $\tilde R_2$ has no direct
couplings to $\ell$. The effective interaction term of $\tau \to \ell \nu_\tau
\bar \nu_\ell$ then reads $\tfrac{-g^2}{2 m_W^2} (\bar \nu_\tau \gamma_
\mu P_L \tau)(\bar \ell \gamma^\mu P_L \ell)
  [1+ \delta_W^{(\tau)}+ \delta_W^{(\ell)}+ \delta^\mrm{box}_{\tau \ell\nu\nu}]$, with
  \begin{equation}
   \delta^\mrm{box}_{\tau \ell\nu\nu} = \frac{N_c}{128\pi^2} \frac{v^2}{m_{S_3}^2}
   \left[(y^\dagger y)^2_{\ell \tau} + 4(y^\dagger y)_{\tau\tau}
     (y^\dagger y)_{\ell\ell} \right].
  \end{equation}
As it has been pointed out recently in the
literature~\cite{Pich:2013lsa,Feruglio:2016gvd,Feruglio:2017rjo} the LFU observable
$R_\tau^{\tau/\ell}$, defined as a ratio $\br(\tau \to \ell \nu\nu)/\br(\mu \to
  e\nu\nu)$, and normalized to the SM prediction of this ratio, is
very sensitive to models modifying couplings of the $\tau$
lepton. Experimentally, $R_\tau^{\tau/\mu} = 1.0022\pm 0.0030$,
$R_\tau^{\tau/e} = 1.0060\pm 0.0030$, while in the present model the
leading interference terms shift the ratios as
\begin{equation}
  \label{eq:RFeruglio}
  R_\tau^{\tau/e} = 1+ 2 \re \left(\delta_W^{(\tau)} - \delta_W^{(\mu)}\right),\qquad
  R_\tau^{\tau/\mu} = 1+ 2 \re \left(\delta_W^{(\tau)}
    +\delta^\mrm{box}_{\tau \mu\nu\nu} \right).
\end{equation}

\subsubsection{Semileptonic decays of $D$ and $t$}

We have checked the effect of $S_3$ on the leptonic charm meson decays $D_s \to \ell \nu$.
Using the bounds from kaon LFU observables presented above we find that
the $S_3$ correction to the $D \to \mu \nu$ width is below $1\%$, while
the experimental uncertainty of $D_s \to \tau \nu$ is $4\%$ and can easily
accommodate $|y_{s\tau}| \lesssim 1.2 (m_{S_3}/\mrm{TeV})$ without even
taking into account the uncertainty in the decay constant
$f_{D_s}$. For the semileptonic top decay process among the third
generation fermions, $t \to b \tau^+ \nu$, the correction is also
below the current sensitivity~\cite{Aaltonen:2014hua}.

\subsection{LFV and neutral currents}

\subsubsection{$\tau \to \mu \gamma$}
Current bound $\br(\tau \to\mu \gamma) \le 4.4\times 10^{-8}$ has been
determined by the BABAR collaboration~\cite{Aubert:2009ag}. The $S_3$ LQ contributes to the $\tau \to \mu \gamma$ amplitude
 via $b$ and $s$ quarks and $S_3^{4/3}$ in the loop and also via up
 quarks $u$, $c$, and $t$ mediated by the $S_3^{-1/3}$ component. Using
 the loop functions in the small quark mass limit as in
 Ref.~\cite{Dorsner:2016wpm} we determine
\begin{equation}
  \label{eq:l-lprimegamma}
 \mc{L}_\mrm{eff}^{\tau \to \mu \gamma} = \frac{e}{2} \sigma^{\tau
   \mu}_L \, \bar\mu (i\sigma^{\mu\nu} P_L)\tau\,  F_{\mu\nu},
\end{equation}
where the effective coupling reads
\begin{equation}
\label{eq:tau-sigma}
\sigma^{\tau \mu}_L = \frac{3 m_\tau}{64 \pi^2 m_{S_3}^2}
\left[5y_{s\mu} y_{s\tau}^* + y_{b\mu} y_{b\tau}^*\right].
\end{equation}

\subsubsection{$Z \to \mu \tau$ and $\tau \to 3\mu$}
At loop level, $S_3$ and $\tilde R_2$ modify the $Z \to f_1 \bar f_2$
decay widths which were precisely measured at LEP-2. The largest
effects in presented LQ model are expected for third generation final states both in flavor conserving
decays, as in $Z\to \tau^+ \tau^-$, which has been shown to have only
weak constraining power in Ref.~\cite{Dorsner:2013tla}, as well as in LFV modes $Z \to
\tau^\pm \mu^\mp$. The latter decay happens due to penguin diagrams with
$S_3$ as well as 1-particle reducible diagrams
and is suppressed by a loop factor and small
ratio $x = m_Z^2/m_{S_3}^2$, in which we expand to leading order:
\begin{equation}
  \label{eq:Zmutau}
  \Gamma_{Z\to \tau^\mp \mu^\pm} = \frac{\sqrt{2} G_F
    m_Z^3}{3\pi}\,\left|\frac{N_c}{288\pi^2} x (2-3 \cos^2\theta_W -3\log
    x+3\pi i) \right|^2\, \left(|y_{s\mu} y_{s\tau}|^2 + y_{b\mu}
    y_{b\tau}|^2 \right).
\end{equation}
We have checked that $\br(Z \to \mu\tau)$ is well below the current experimental
bound at $10^{-5}$. Compared to the closely related $\tau \to \mu
\gamma$ decay, this channel is less stringently constrained and thus
we do not include it in the fit. On the other hand, the $\br(\tau \to 3 \mu) <
2.1\E{-8}$ at 90\% C.L.~\cite{Olive:2016xmw}, and can be mediated by
the above mentioned LFV $Z$ vertex or via box diagram containing $S_3$
and quarks. They are both encompassed in the
low-energy effective Lagrangian:
\begin{equation}
  \label{eq:tau3mu}
  \begin{split}
  \mc{L}_{\tau \to 3\mu} &= \frac{-N_c (y^\dagger y)_{\mu\tau}}{(4\pi)^2 m_{S_3}^2}
  \left[ (y^\dagger y)_{\mu\mu} +
    \frac{\sqrt{2}}{9} G_F m_W^2 (2-3 \cos^2\theta_W -3\log
    x-3\pi i)\right]\, (\bar \mu \gamma^\mu P_L \tau) (\bar \mu
  \gamma_\mu P_L \mu) \\
& \phantom{=} \frac{-N_c (y^\dagger y)_{\mu\tau}}{(4\pi)^2 m_{S_3}^2}
    \frac{2\sqrt{2}}{9} G_F m_Z^2 \sin^2 \theta_W (2-3 \cos^2\theta_W -3\log
    x-3\pi i)\, (\bar \mu \gamma^\mu P_L \tau) (\bar \mu
  \gamma_\mu P_R \mu),
  \end{split}
\end{equation}
where, again, $x = m_Z^2/m_{S_3}^2$.
The mixed chirality stems from the $Z$ coupling to $\bar \mu_R \mu_R$.  In the limit
of $m_\mu/m_\tau \to 0$ the two terms above do not interfere. We
notice that when all couplings are $\sim 1$ and $m_{S_3} =
1\e{TeV}$ then the $\br(\tau \to 3\mu)$ is in the ballpark of current
experimental upper bound. As will be shown in Sec.~\ref{sec:fit}, realistic values of the Yukawas result in
much smaller contribution to this channel, and that is why we omit
this channel from the fit.

\subsubsection{$(g-2)_\mu$}
The difference between the experimental value and the one predicted by
the SM is
$\delta a_\mu = a_\mu^\mrm{exp} - a_\mu^\mrm{SM}= (2.8\pm 0.9)\times
10^{-9}$~\cite{Olive:2016xmw}. Following~\cite{Dorsner:2016wpm} and using the
Lagrangian of Eq.~\eqref{eq:S3}, we derive the contribution of $S_3$ to the
muon anomalous magnetic moment:
\begin{equation}
\label{eq:muon}
\delta a_\mu^{S_3} = -\frac{3 m_\mu^2}{(32 \pi^2m_{S_3}^2} \left(|y_{s\mu}|^2 + |y_{b\mu}|^2\right).
\end{equation}
Since the above contribution has wrong sign with respect to the
experimental pull individual
Yukawa couplings of $S_3$ to the $\mu$ should be small. Notice that
the contribution of $\tilde R_2$ to $(g-2)_\mu$ is greatly
suppressed and vanishes at $m_{s,b}/m_{\tilde R_2} \to 0$~\cite{Dorsner:2016wpm,Queiroz:2014pra}.

\subsubsection{$B \to K \mu  \tau$ decays}
The lepton flavor violation can be induced by the LQ presence at tree
level in $B \to K \mu \tau$ and also in decays of bottomonium to
$\tau \mu$. As noticed in~\cite{Fajfer:2015ycq} the latter process has
been constrained at the level of $10^{-6}$ however these bounds are
not competitive with the bound
$\br(B^- \to K^- \mu^\pm \tau^\mp) < 4.8\E{-5}$ at 90\%
C.L.~\cite{Lees:2012zz}. This inclusive mode is sensitive to both
$y_{s\tau} y_{b \mu}$ and $y_{s\mu} y_{b\tau}$ as $\br(B^- \to
K^- \mu^\pm \tau^\mp) = 8.6\E{-3} [(y_{s\tau} y_{b \mu})^2 + (y_{s\mu}
y_{b\tau})^2]$ when the form factors of Ref.~\cite{Bailey:2015dka} are
used. The constraint then reads
\begin{equation}
  \label{eq:BKtaumu}
\sqrt{(y_{b \tau} y_{s\mu})^2 + (y_{b \mu} y_{s\tau})^2}  \lesssim 0.075 (m_{S_3}/\mrm{TeV})^2.  
\end{equation}

\subsubsection{$B_s$--$\bar B_s$ mixing frequency}
\begin{figure}[!htbp]
  \centering
  \begin{tabular}{lcccr}
    \includegraphics[scale=0.68]{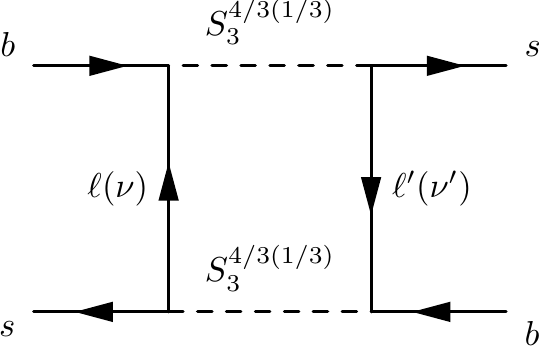}
    & \hspace{1cm} & \includegraphics[scale=0.68]{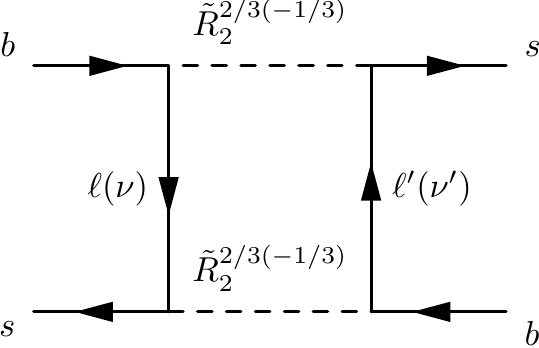} &\hspace{1cm}  &\includegraphics[scale=0.68]{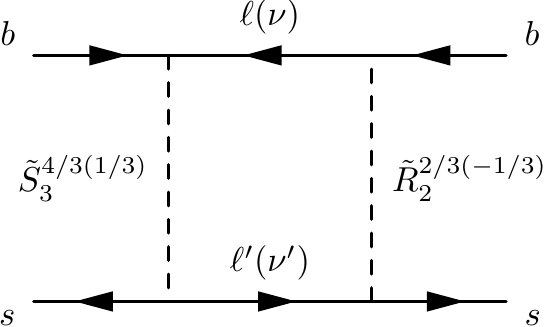}
  \end{tabular}
  \caption{Three types of box-diagrams with $S_3$ and $\tilde R_2$ contributing to
    $\Delta m_s$.}
  \label{fig:BsMix}
\end{figure}

Despite being a loop observable in the LQ scenarios, the $B_s$ meson
mixing frequency is one of the most important constraints in our
particular setup where the product of $S_3$ Yukawas
$y_{b\tau} y_{s\tau}$ is large. This product alone would lead to
uncomfortably large effect in the $B_s$--$\bar B_s$ oscillation frequency
$\Delta m_s$. However, there is an additional box amplitude due to
$\tilde R_2$ as well as an amplitude with both $S_3$ and $\tilde R_2$
propagating in the box, as shown in Fig.~\ref{fig:BsMix}.
Amplitudes that correspond to the first and second diagram in Fig.~\ref{fig:BsMix}
can be found in Refs.~\cite{Dorsner:2016wpm, Becirevic:2016yqi} and contribute to operators
$C_1$ and $\tilde C_1$ of the effective Hamiltonian, respectively:
\begin{equation}
  \label{eq:MixH}
  \mc{H}_{\Delta m_s} = (C_1^\mrm{SM} + C_1^{S_3})\,(\bar s_L \gamma^\nu b_L)^2 +
  \tilde C_1^{\tilde R_2} \, (\bar s_R \gamma^\nu b_R)^2 + C_4^{S_3
    \tilde R_2} \,(\bar s_R b_L)(\bar s_L b_R) +C_5^{S_3
    \tilde R_2} \,(\bar s_R^\alpha b_L^\beta)(\bar s_L^\beta b_R^\alpha).
\end{equation}
The third diagram in Fig.~\ref{fig:BsMix} in which both LQs are
present, but couple with opposite chirality
to the fermions, contributes to the Wilson
coefficient $C_5$. There the color indices $\alpha$ and $\beta$ 
are summed across $\Delta B = 1$ currents. The box diagrams in Fig.~\ref{fig:BsMix}
are well approximated using a limit of massless virtual leptons and
match onto the effective Hamiltonian at scale $\Lambda =
\mc{O}(m_{S_3}) \sim 1\e{TeV}$:
\begin{equation}
  \label{eq:MixMatch}
  \begin{split}
  C_1^\mrm{SM}(m_t) &= \frac{m_W^2 S_0 (x_t) (V_{tb} V_{ts}^*)^2}{8\pi^2
    v^4},\\
C_1^{S_3} (\Lambda) &= \frac{3(yy^\dagger)^2_{bs}}{128\pi^2 m_{S_3}^2},\\
\tilde C_1^{\tilde R_2} (\Lambda)&= \frac{(\tilde y \tilde y^\dagger)^2_{sb}}{64 \pi^2
  m_{R_2}^2},\\
C_4^{S_3\tilde R_2}(\Lambda) &= 0,\\
C_5^{S_3\tilde R_2}(\Lambda) &= \frac{(y \tilde y^\dagger)_{bb}(\tilde y
  y^\dagger)_{ss}}{16 \pi^2} \frac{\log m_{S_3}^2/m_{R_2}^2}{m_{S_3}^2-m_{R_2}^2}.
  \end{split}
\end{equation}
Evaluation of hadronic matrix elements for $B_s$--$\bar B_s$ mixing
is performed at the scale $\mu = \bar m_b(\bar m_b) =
4.2\e{GeV}$. Utilizing parameterization in terms of bag parameters as
in Ref.~\cite{Gabbiani:1996hi}, we find for the oscillation frequency
\begin{equation}
  \label{eq:DeltaMs}
  \begin{split}   
  \Delta m_s &= \frac{2}{3} m_{B_s} f_{B_s}^2 B_{B_s}^{(1)}(\mu)
  \left|C_1^\mrm{SM}(\mu)\right|  \\
& \quad  \times \left| 1+\left[\frac{C_1^{S_3}+\tilde
      C_1^{\tilde R_2}}{C_1^\mrm{SM}}\right]_\mu  +
  \frac{1}{2}\left[\left(\frac{m_{B_s}}{\bar m_b(\bar m_b)+\bar
        m_s(\bar m_b)}\right)^2 +\frac{3}{2}\right]
  \left[\frac{B_{B_s}^{(5)}}{B_{B_s}^{(1)}} \frac{C_5^{\tilde R_2
      S_3}}{C_1^\mrm{SM}}\right]_\mu
\right. \\
&\quad\qquad + \left.
\frac{3}{2}\left[\left(\frac{m_{B_s}}{\bar m_b(\bar m_b)+\bar
        m_s(\bar m_b)}\right)^2 +\frac{1}{6}\right]
  \left[\frac{B_{B_s}^{(4)}}{B_{B_s}^{(1)}} \frac{C_4^{\tilde R_2
      S_3}}{C_1^\mrm{SM}}\right]_\mu
\right|.
  \end{split}
\end{equation}
For the SM prediction we use the perturbative QCD renormalization
at next-to-leading order whose effect is subsumed in
$\eta_{2B} = 0.55(1)$~\cite{Buras:1990fn}. The non-perturbative
parameters and perturbative RG running effects of $C_1$ are combined into
a scale-invariant combination,
\begin{equation}
  f_{B_s}^2 B_{B_s}^{(1)}(\mu) C_1^\mrm{SM}(\mu) =
   f_{B_s}^2 \hat B_{B_s}^{(1)} \eta_{2B} C_1^\mrm{SM}(m_t),
\end{equation}
where the value of renormalization-group invariant bag parameter is
taken from the QCD lattice simulation with three dynamical
quarks~\cite{Bazavov:2016nty}:
$f_{B_s}^2\hat B_{B_s}^{(1)} = 0.0754(46)(15)\e{GeV}^2$~\footnote{We
  prefer to use the results of Ref.~\cite{Bazavov:2016nty} that
  include bag parameters for the whole operator basis. However, for
  $B_{B_s}^{(1)}$ we have found good agreement with the FLAG average
  of 2+1 dynamical simulations,
  $f_{B_s}^2\hat B_{B_s}^{(1)} =
  0.0729(86)$~\cite{Aoki:2016frl}.}. First number in the brackets
represents statistical and systematic error, apart from systematic
error due to omission of dynamical charm-quark, which is shown in the
second bracket. The SM prediction is then $\Delta m_s^\mrm{SM} =
(19.6\pm 1.6)\e{ps}^{-1}$.
 For the LQ contributions in Eq.~\eqref{eq:DeltaMs} we
use the values of $B_{B_s}^{(i)}(\mu)$ from
Ref.~\cite{Bazavov:2016nty}. For the multiplicative renormalization of
coefficients $C_1^{S_3}$ and $\tilde C_1^{\tilde R_2}$
we neglect the running from $\Lambda$ to $m_t$, such that running
effect to low scale is the same as in the SM, whereas for
$C_{4,5}^{\tilde R_2 S_3}$ we use the leading order
mixing~\cite{Fajfer:2006av} to find
$C_{4}^{\tilde R_2 S_3}(\mu) = 0.61 C_{5}^{\tilde R_2 S_3}(\Lambda)$,
$C_{5}^{\tilde R_2 S_3}(\mu) = 0.88 C_{5}^{\tilde R_2
  S_3}(\Lambda)$. For the ratios of bag parameters we use central
values to find $B_{B_s}^{(5)}(\mu)/B_{B_s}^{(1)}(\mu) = 0.99$,
$B_{B_s}^{(4)}(\mu)/B_{B_s}^{(1)}(\mu) =
1.07$~\cite{Bazavov:2016nty}. Note that in this case the experimental
value $\Delta m_s^\mrm{exp} = (17.757 \pm 0.021)\e{ps}^{-1}$ has
negligible uncertainty~\cite{Olive:2016xmw}.

\subsubsection{$B \to K^{(*)} \nu \bar \nu$}
The $B \to K^{(*)} \nu \bar\nu$ decay offers an excellent probe of the
lepton flavor conserving as well as lepton flavor violating
combination of the LQ couplings. Following~\cite{Fajfer:2015ycq} and
with the help of notation in
Refs.~\cite{Altmannshofer:2009ma,Buras:2014fpa,Becirevic:2015asa}, we
write the effective Lagrangian:
\begin{equation}
\mc{L}_\mrm{eff}^{b\to s \bar \nu \nu} = \frac{G_F \alpha }{\pi \sqrt{2}} V_{tb} V_{ts}^*
\left (\bar s \gamma_\mu [C_L^{ij} P_L + C_R^{ij} P_R] b\right)(\bar \nu_i \gamma^\mu (1-\gamma_5) \nu_j).
\end{equation}
In the SM we have a contribution for each pair of neutrinos and therefore 
$C_L^{\mrm{SM},ij} = C_L^\mrm{SM}\delta_{ij}$ where $C_L^\mrm{SM} =
-6.38 \pm 0.06$~\cite{Altmannshofer:2009ma}. The respective contributions of $S_3$ and $\tilde R_2$ to the left-
and right-handed operators are~\cite{Dorsner:2016wpm}:
\begin{equation}
  \label{eq:BKnunu}
  C_L^{S_3,ij} = \frac{\pi v^2}{2\alpha V_{tb} V_{ts}^* m_{S_3}^2} y_{bj}
  y^*_{si},\qquad   C_R^{\tilde R_2,ij} = -\frac{\pi v^2}{2\alpha V_{tb} V_{ts}^*
    m_{\tilde R_2}^2} \tilde y_{sj}  \tilde y^*_{bi}.
\end{equation}
While the amplitude of $B\to K \bar \nu \nu$ depends only on the
vectorial part of Wilson coefficients~\eqref{eq:BKnunu}, the $B \to
K^* \bar \nu \nu$ amplitude is also sensitive to axial current, and 
the two decay modes constrain the right-handed Wilson coefficient
differently. We follow Ref.~\cite{Buras:2014fpa} and introduce
\begin{equation}
  \label{eq:epseta}
  \epsilon_{ij} = \frac{\sqrt{|C_L^\mrm{SM} \delta_{ij} +
      C_L^{S_3,ij}|^2 + |C_R^{\tilde R_2,ij}|^2}}{|C_L^\mrm{SM}|},
  \quad \eta_{ij} = \frac{-\re\left[(C_L^\mrm{SM} \delta_{ij} +
      C_L^{S_3,ij}) C_R^{\tilde R_2,ij\ast} \right]}{|C_L^\mrm{SM} \delta_{ij} +
      C_L^{S_3,ij}|^2+|C_R^{\tilde R_2,ij}|^2}.
\end{equation}
Then the SM-normalized branching fractions are
\begin{equation}
  \label{eq:Rnunu}
  \begin{split}
      R_{\nu\nu} &= \frac{\br(B\to K\bar\nu \nu)}{\br(B\to K\bar\nu
    \nu)_\mrm{SM}} = \frac{1}{3} \sum_{ij}
  (1-2\eta_{ij})\epsilon_{ij}^2,\\
R^\ast_{\nu\nu} &= \frac{\br(B\to K^*\bar\nu \nu)}{\br(B\to K^*\bar\nu
    \nu)_\mrm{SM}} = \frac{1}{3} \sum_{ij}
  (1+ \kappa_\eta \eta_{ij})\epsilon_{ij}^2,
  \end{split}
\end{equation}
where $\kappa_\eta= 1.34\pm 0.04$~\cite{Buras:2014fpa}.
Among the possible final states, we will take the two strongest bounds
on $R_{\nu\nu}^{(\ast)}$ determined by the Belle experiment,
$\br(B \to K^* \nu \bar \nu) < 2.7 \times 10^{-5}$ and
$\br(B \to K \nu \bar \nu) < 1.6 \times 10^{-5}$ which translate to
$R_{\nu\nu}^* < 2.7$ and $R_{\nu\nu} < 3.9$, both at $90\%$
C.L.~\cite{Grygier:2017tzo}.

\subsubsection{$b\bar b \to \mu^+ \mu^-$ scattering}
The measurements of $\mu^+ \mu^-$ spectra at high invariant mass
$m_{\mu\mu}$ are sensitive to large couplings $y_{s\mu}$ or
$y_{b\mu}$. The relevant channel in our case is $b\bar b \to \mu^+
\mu^-$ which directly limits $y_{b\mu}$. If we assume that effective
dim-6 operator description is a good approximation to the $t$-channel
$S_3$ exchange at LHC energy, then we can use a $1\,\sigma$ bound derived in
Ref.~\cite{Greljo:2017vvb}
\begin{equation}
  \label{eq:bbmumu}
  y_{b\mu}^2 < 0.30 (m_{S_3}/\mrm{TeV})^2 .
\end{equation}

\subsubsection{$D$ decays}
The weak triplet nature of $S_3$ implies couplings only to the
weak doublets of quarks and leptons, and thus corrections to the charged
current processes only rescale the SM charged current contributions.
The dominant modification of $V_{cs}$ element associated
with semi-muonic decays follows from Eq.~\eqref{eq:gUDlnu}:
\begin{equation}
  \label{eq:charmSL}
V_{cs} \to V_{cs}  -\frac{v^2}{4 m_{S_3}^2} (y_{s\mu} + V_{cb}^*
y_{b\mu} )  y_{s\mu}.   \qquad (\textrm{for processes with }\mu \bar \nu_\mu).
\end{equation}
Assuming that the CKM-suppressed $y_{b\mu}$ term can be neglected in
Eq.~\eqref{eq:charmSL} and using the fact that current precision on
the semileptonically determined $V_{cs}$ reaches 1 per-mille~\cite{Olive:2016xmw}, we find
$y_{s\mu} \lesssim 0.3 (m_{S_3}/\mrm{TeV})$.

Rare charm decays with two leptons, e.g. $D^0 \to \mu^+ \mu^-$ and $D
\to M \mu^+ \mu^-$, are most constraining at the moment (for dineutrino
modes cf.~\cite{deBoer:2015boa}), where $M$ can be a pseudoscalar or a
vector meson. The effective Wilson coefficient of the left-handed current, $C_9 = -C_{10}
\approx (V_{us} \pi v^2)/(\alpha V_{ub} V_{cb}^* m_{S_{3}}^2)
y_{s\mu}^2$ can be compared to the bounds, $|C_9|, |C_{10}| \lesssim
1.0/|V_{ub} V_{cb}|$, obtained in Ref.~\cite{Fajfer:2015mia}. We learn that
the ensuing bound $y_{s\mu} \lesssim 0.5 (m_{S_3}/\mrm{TeV})$ from rare
decays is weaker than the abovementioned bound from semileptonic decays.

\begin{figure}[!bhtp]
  \centering
  \begin{tabular}{lcr}
   \includegraphics[scale=0.37]{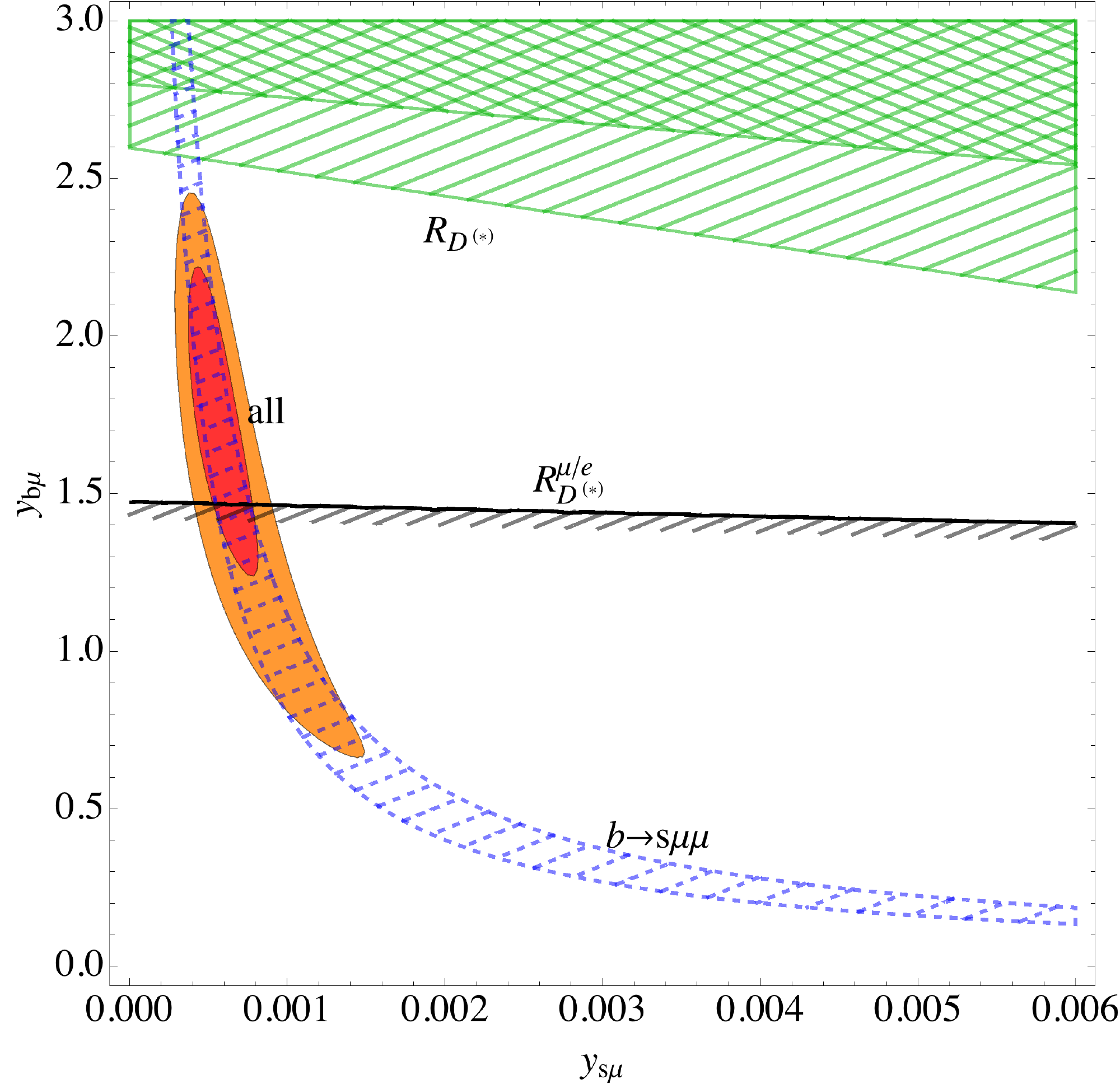}  && \includegraphics[scale=0.37]{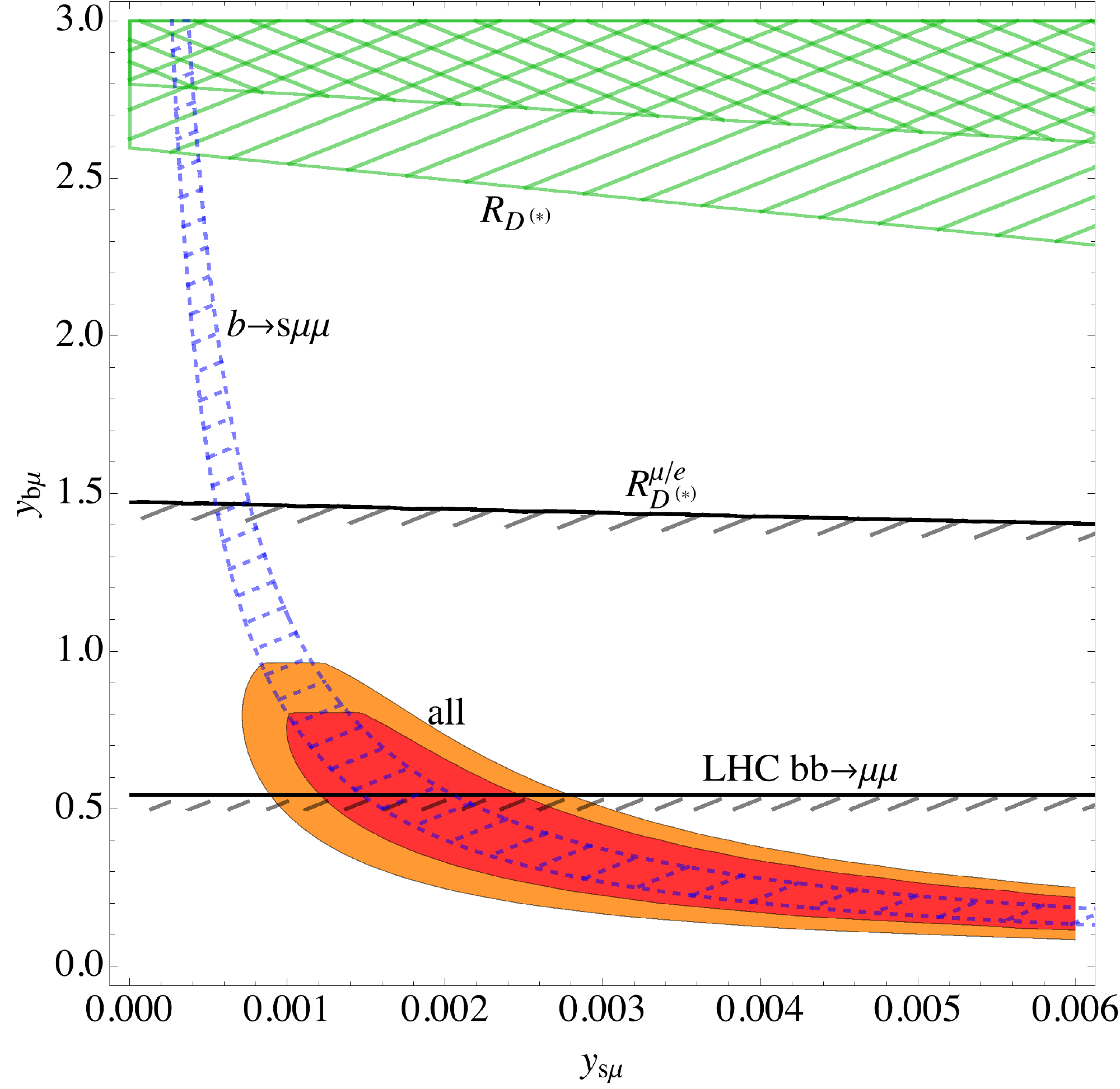}
  \end{tabular}
  \caption{\label{fig:sce1fig1}Left panel: $R_{D^{(*)}}$ is resolved in
    hatched ($2\,\sigma$) and doubly hatched ($1\,\sigma$) regions,
   whereas the $b\to s \mu\mu$ puzzle is resolved in dashed-hatched region at $1\,\sigma$.
   Region below the black line with a hatching is in $1\,\sigma$
   agreement with $R_{D^{(*)}}^{\mu/e}$. No LHC constraint on
   $y_{b\mu}$ is considered. Right panel: Same as left panel apart from
   inclusion of constraint on $y_{b\mu}$ from LHC.
 Red and orange regions in both graphs denote $1\,\sigma$
    and $2\,\sigma$ results of the fit.}
\end{figure}
\section{Flavor couplings}
\label{sec:fit}
In this section we study three scenarios differing in the number
of variable Yukawas. For each scenario we report a minimum of $\chi^2$
function, which is a sum of terms corresponding to all observables
discussed in the preceding sections.
We also report 
$1\,\sigma$ regions for the interesting two-dimensional projections of
parameter space. While performing these fits we limit all free Yukawa
couplings to be smaller than $3$. Introduction of this artificial
cut-off is guided by the constraints posed by the LHC searches,
discussed in Sec.~\ref{sec:collider}.  The SM point has $\chi^2 =
71.6$ and serves as a reference value to which $\chi^2$ of the
three fits are compared.

\subsection{$S_3$ coupled to muons (2 parameters)}
In this scenario we consider only the effect of $S_3$ with
non-zero muonic couplings:
\begin{equation}
y=   \begin{pmatrix}
  0 & 0 & 0\\
  0 & y_{s \mu} & 0\\
  0 & y_{b \mu} & 0
\end{pmatrix}.
\end{equation}
We set $m_{S_3} = 1\e{TeV}$ and for the moment ignore the direct LHC
constraint on $y_{b\mu}$ spelled out in Eq.~\eqref{eq:bbmumu}. In this
case the best fit point has $\chi^2 = 34.7$ reached at
$y_{s\mu} = 5\E{-4}$ and $y_{b\mu} = 1.8$. The $R_{D^{(*)}}$ puzzle
can be addressed by lowering $\br(B \to D^{(*)}\mu\nu)$ which requires
large $y_{b\mu}$ coupling as seen in Eq.~\eqref{eq:RDcons}. The $1\sigma$
and $2\sigma$ regions of the fit are shown in
Fig.~\ref{fig:sce1fig1}. Left panel in Fig.~\ref{fig:sce1fig1} exposes
tension between $R_{D^{(*)}}$ ($2.8\,\sigma$ pull) and
$R_{D^{(*)}}^{\mu/e}$ ($1.8\,\sigma$ pull) which is even more
exacerbated when we include the direct constraints on $y_{b\mu}$ from
LHC (right panel of Fig.~\ref{fig:sce1fig1}). The latter scenario with all constraints included has
$\chi^2=42.4$ at point $(y_{s\mu}, y_{b\mu}) = \pm(2\E{-3}, 0.46)$ which
corresponds to the $5.0\,\sigma$ pull of the SM hypothesis. One can
observe in the right panel in Fig.~\ref{fig:sce1fig1} that in this
case the preferred region is drawn further away from $R_{D^{(*)}}$.
The results indicate that $R_{D^{(*)}}$ cannot be explained by
omitting couplings to $\tau$. Detailed results on the pulls are given
in the third column of Tab.~\ref{tab:fitresults}.
\begin{table}[!htbp]
  \centering
  \begin{tabular}{|c|c|c|c|c|}
\hline
    &SM & $m_{S_3} = 1\e{TeV}$ & $m_{S_3} = 1.0/1.5\e{TeV}$ & Eq.\\
&  & $(y_{s\mu}, y_{b\mu})$ & $(y_{s\mu}, y_{b\mu},y_{s\tau},
                              y_{b\tau})$&\\
&  & w.o./w. Eq.~\eqref{eq:bbmumu} & &\\\hline
$\chi^2$    & $71.6$ & $34.7/42.4$  & 36.8/38.0 &\\\hline\hline
$b\to s \ell^+ \ell^-$ & 5.4 & 0.0/0.0& 0.0/0.0&\eqref{eq:bsmumuCons} \\\hline
$R_{D^{(*)}}$ & 4.5 & 2.8/4.4& 4.0/4.2&\eqref{eq:RDcons}\\\hline
$(g-2)_\mu$ & 3.1 & 3.5/3.1& 3.1/3.1&\eqref{eq:muon} \\\hline
$R^K_{\tau/\mu}$ &2.0 & 2.0/2.0& 0.3/0.3&\eqref{eq:Rtaumu} \\\hline
$R^{\tau/e}_\tau$ &2.0 & 1.6/2.0& 2.1/2.1&\eqref{eq:RFeruglio} \\\hline
$\br(B \to \tau \nu)$ &1.2 & 1.2/1.2& 1.1/1.2& \eqref{eq:Btaunu} \\\hline
$\Delta m_s$ &1.1 & 1.1/1.1& 1.6/1.6&\eqref{eq:DeltaMs} \\\hline
$R^K_{e/\mu}$ &1.1 & 1.1/1.1& 1.1/1.1&\eqref{eq:RKemu}\\\hline
$R^{\tau/\mu}_\tau$ & 0.7 & 0.7/0.7& 0.8/0.8&\eqref{eq:RFeruglio} \\\hline
$R_{D^{(*)}}^{\mu/e}$ & 0.5 & 1.8/0.4& 0.5/0.5&\eqref{eq:RDmue} \\\hline
$R_{\nu\nu}$ & 0.5 & 0.6/0.6& 0.8/0.6&\eqref{eq:Rnunu} \\\hline
$bb\to \mu\mu$ & 0.0 & $\,-\,$/0.7 & 0.0/0.0&\eqref{eq:bbmumu}   \\\hline
$\br(\tau \to \mu \gamma)$ & 0.0 & 0.0/0.0&
                                            0.4/0.3&\eqref{eq:l-lprimegamma}\\\hline
$\br(B \to K \tau \mu)$ & 0.0 & 0.0/0.0&
                                            0.3/0.3&\eqref{eq:BKtaumu}\\\hline
  \end{tabular}
  \caption{Observables that enter the global fit with their pulls in $\sigma$ in
    the SM and $S_3$ scenarios. Third column represents the case
    when $m_{S_3} = 1\e{TeV}$ and only $y_{s\mu},y_{b\mu}$ are
    allowed, without/with taking into account $b\bar b \to \mu\mu$
    constraint. Fourth column represents the fit of the $y_{s\mu}$,
    $y_{b\mu}$, $y_{s\tau}$, $y_{b\tau}$ scenario for $m_{S_3} = 1.0/1.5\e{TeV}$.
 The constraints with negligible pulls are
    not shown in this table.}
  \label{tab:fitresults}
\end{table}

\subsection{$S_3$ coupled to muons and taus (4 parameters)}
\label{subsection:fourY}
\begin{figure}[!htbp]
  \centering\includegraphics[scale=0.5]{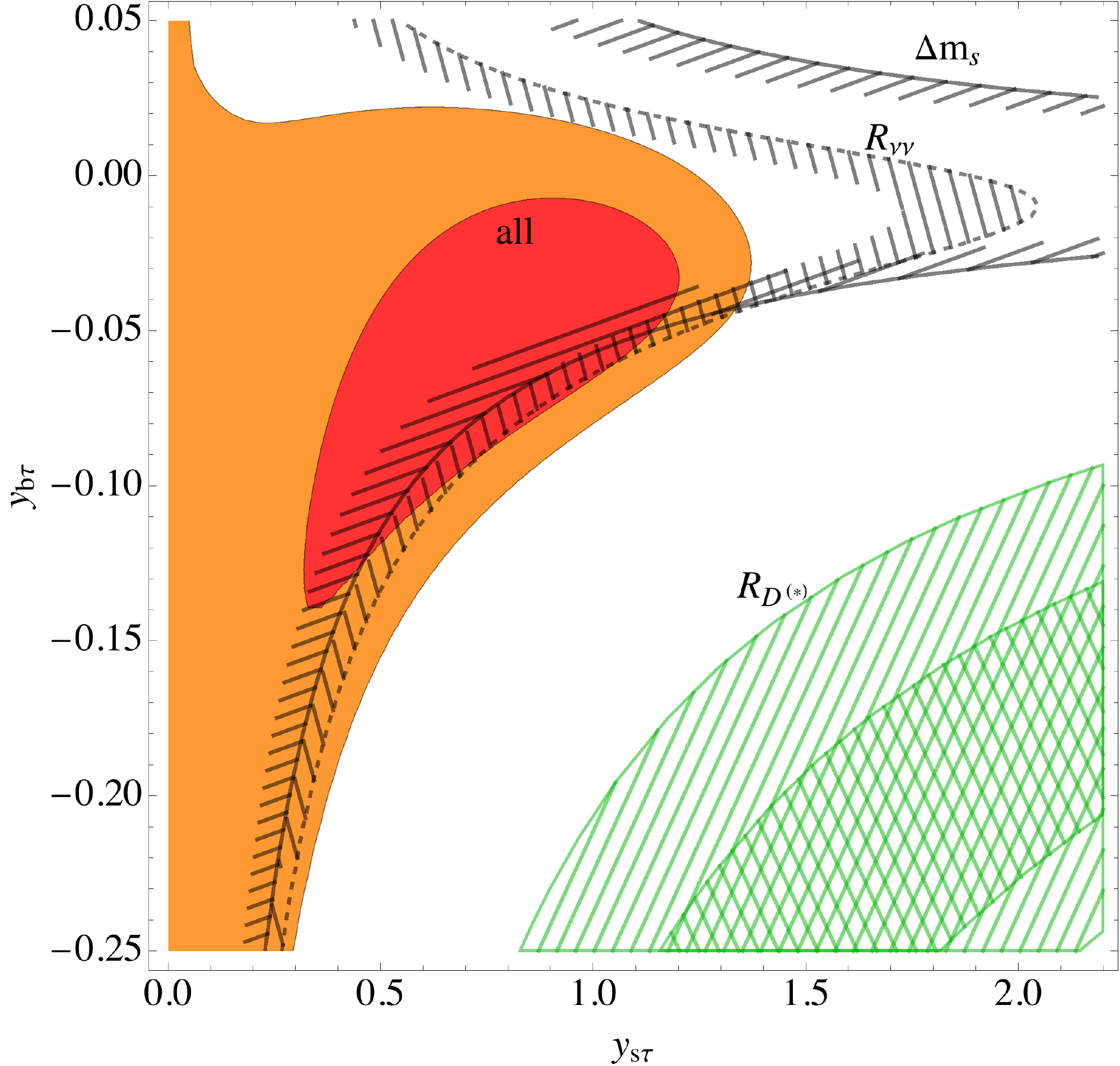}
  \caption{\label{fig:scen2} Fit for the $m_{S_3} = 1\e{TeV}$ scenario
    with four free couplings. $R_{D^{(*)}}$ is resolved within hatched
    ($2\,\sigma$) and doubly hatched ($1\,\sigma$) regions.  Region to
    the left of the dashed line (hatched) is in $1\,\sigma$ agreement with
    $R_{\nu\nu}$ and $R_{\nu\nu}^*$. $\Delta m_s$ prefers (at
    $2\,\sigma$) a region on the hatched side of full line.  Red and
    orange regions are $1\,\sigma$ and
    $2\,\sigma$ results of the fit.}
\end{figure}
Since the purely muonic couplings are in conflict with $R_{D^{(*)}}$
we allow in addition for tauonic couplings of $S_3$:
\begin{equation}
y=   \begin{pmatrix}
  0 & 0 & 0\\
  0 & y_{s \mu} & y_{s \tau}\\
  0 & y_{b \mu} & y_{b \tau}
\end{pmatrix}.
\end{equation}
In this case both couplings with the muons tend to be small, below $0.1$,
and are relevant only in $b\to s \mu\mu$, whereas the couplings to $\tau$ are $\sim 1$ in order to enhance
$R_{D^{(*)}}$. For $m_{S_3} = 1\e{TeV}$ we find that the minimal
$\chi^2$ of this scenario
with 4 degrees of freedom is $36.8$ reached at $(y_{s\mu},y_{b\mu}, y_{s\tau},y_{b\tau}) =
(0.047, 0.020, 0.87, -0.048)$\footnote{The fit is approximately invariant with
  respect to the overall sign of the muonic or tauonic couplings which
  implies a fourfold degeneracy.} which makes the SM point excluded
at $5.0\,\sigma$ (pull). In Fig.~\ref{fig:scen2} the fit in the tauonic
couplings' plane shows how the optimal region is still far from
the central value of $R_{D^{(*)}}$, mostly due to $R_{\nu\nu}$ and
$\Delta m_s$, which do not allow for large products of $y_{b\tau}
y_{s\tau}$. Pulls of individual observables for $m_{S_3} =
1.0/1.5\e{TeV}$ are presented in the fourth
column of Table~\ref{tab:fitresults}.

\subsection{$S_3$ and $\tilde R_2$ (6 parameters)}
\label{subsection:sixY}
In order to relax the tension in the $y_{s\tau}$--$y_{b\tau}$ plane
between large effect in $R_{D^{(*)}}$ and well constrained
$R_{\nu\nu}^{(*)}$ and $\Delta m_s$, we could invoke a light
$\tilde R_2$ with couplings to $\tau$. We consider a case
$m_{S_3} = m_{\tilde R_2} = 1\e{TeV}$ with six free Yukawa couplings
($y_{ij}$ from the previous subsection and
$(\tilde y_{s\tau}, \tilde y_{b\tau})$ pair) to find $\chi^2 = 33.4$
at
$( y_{s\mu}, y_{b\mu} , y_{s\tau}, y_{b\tau})=(0.051, 0.019, 0.86,
-0.069)$, $(\tilde y_{s\tau}, \tilde y_{b\tau}) = (3,
0.0026)$\footnote{Degenerate best-fit points are obtained by flipping
  sign of individual Yukawas in a manner that does not change signs of
  $y_{s\mu} y_{b\mu}$, $y_{s\tau} y_{b\tau}$,  $\tilde y_{s\tau}
  \tilde y_{b\tau}$, and $y_{s\tau} \tilde y_{s\tau}$.} that
represents a $4.9\,\sigma$ pull of the SM.  Most importantly, the tension
in $R_{D^{(*)}}$ is only marginally improved and stands at
$3.7\,\sigma$. The presence of $\tilde R_2$ allows for partial
cancellation in $\Delta m_s$ between large tauonic couplings of $S_3$
and $\tilde R_2$, which is not the case in both $R_{\nu\nu}$ and
$R_{\nu\nu}^*$, where cancellation in one observable necessary
spoils the other (cf.~\eqref{eq:Rnunu}). We thus conclude that light
$\tilde R_2$ with relatively large couplings to the SM fermions cannot
improve substantially the agreement with data. We accordingly assume
that the couplings of $\tilde R_2$, i.e., $\tilde{y}_{ij}$ of
Eq.~\eqref{eq:R2}, are small enough as not to affect flavor
observables. With light $\tilde R_2$ and its Yukawa couplings
sufficiently small to avoid flavor constraints we can still aid gauge
coupling unification and generate viable neutrino masses in the
underlying GUT model as we show in Sec.~\ref{sec:gut}.

\section{Collider constrains}
\label{sec:collider}

In what follows we confront our model, comprising two light LQs, with collider constraints while taking into account the particularities of the flavor structure derived in the previous section. We demonstrate the viability of the proposed model and present bounds from third generation LQ pair production as well as high-mass $\tau\tau$ production searches at the 13\,TeV LHC for current and projected luminosities. We show that a large portion of the relevant parameter space can be covered by the HL-LHC.

\begin{figure}[b!]
\centering
\includegraphics[width=0.45\textwidth]{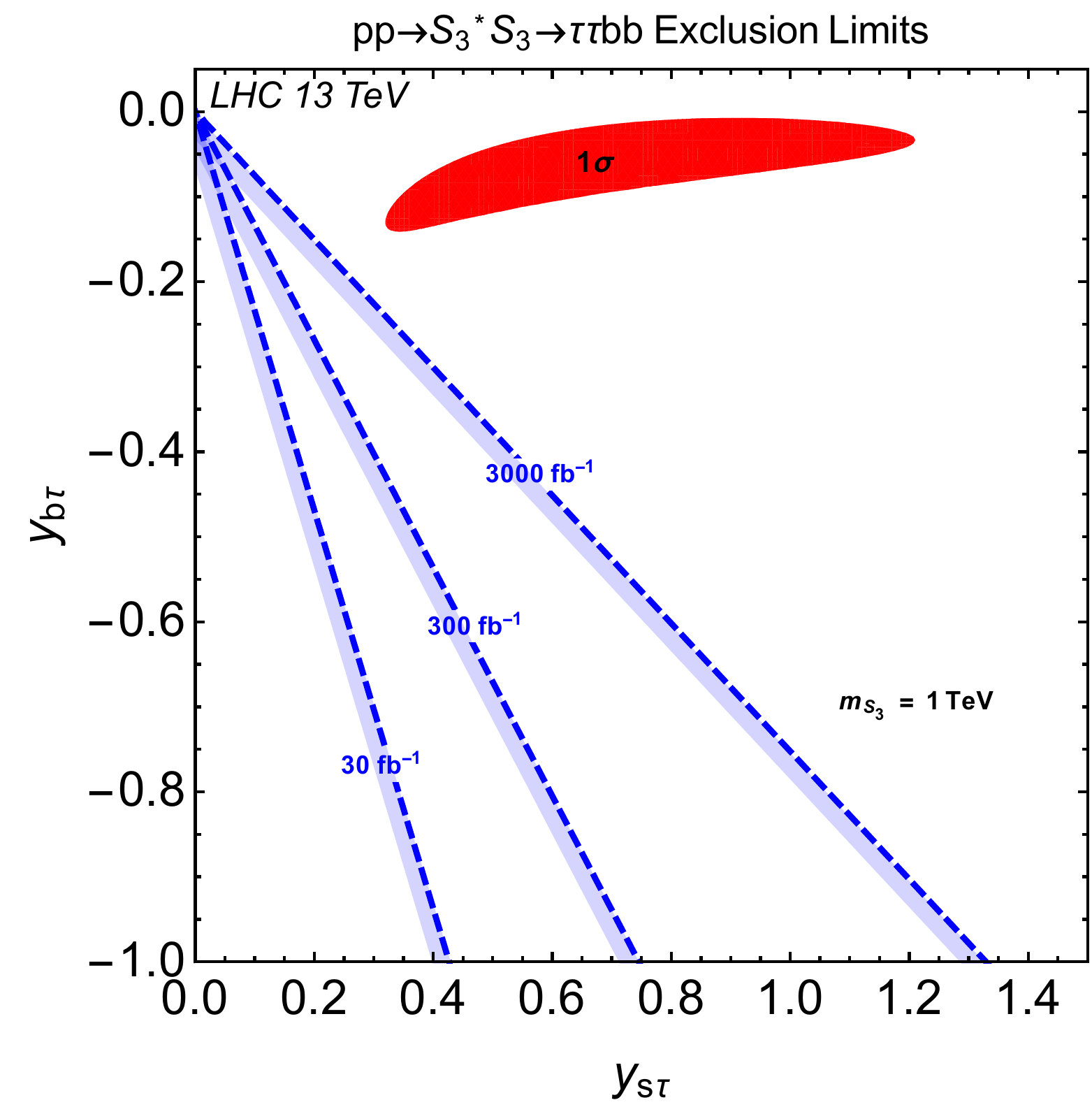}
\includegraphics[width=0.425\textwidth]{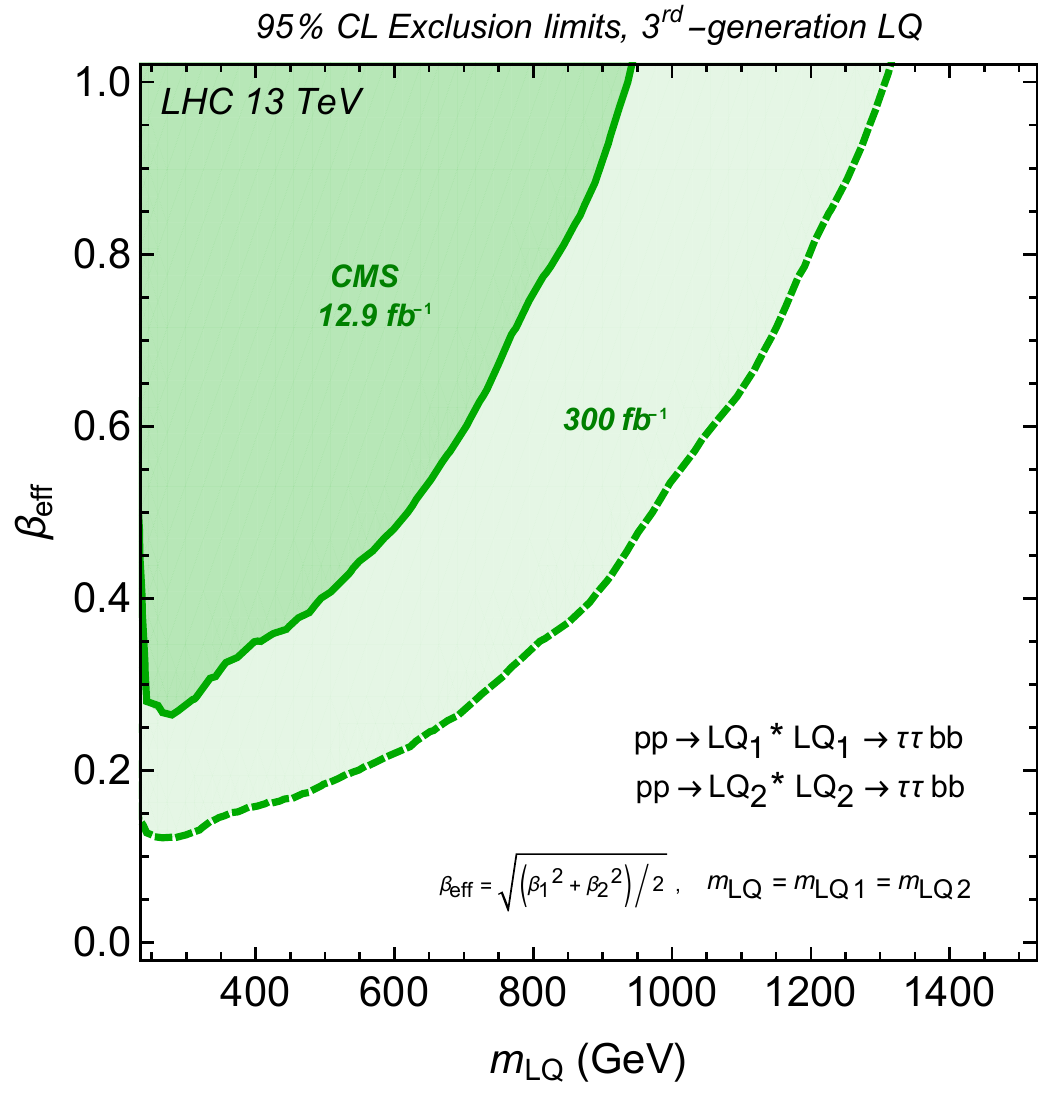}
\caption{\label{fig:2LQrecast} (Left Panel) 95\% C.L. exclusion limits from LQ pair production (dotted blue lines) at different projected LHC luminosities for a 1\,TeV $S_3$ LQ. The red region corresponds to the 1\,$\sigma$ low energy fit. (Right panel) Reinterpretation of the CMS Collaboration exclusion limits for two generic degenerate LQs decaying into $\tau\tau bb$ final state in the $\beta_{\text{eff}}$--$m_\mathrm{LQ}$ plane.}
\end{figure}
\subsection{LQ pair production}
The current best mass limit for LQs that decay to the third-generation leptons has been recently reported by the CMS Collaboration while searching for a pair of QCD produced LQs decaying into the $\tau^+\tau^- b\bar b$ channel~\cite{Sirunyan:2017yrk}. This search excludes an LQ with mass bellow $850$\,GeV (550\,GeV) for a branching ratio (BR) of $\beta=1$ ($\beta=0.5$). This search can set limits on the parameter space of our model via $pp\!\to\! S_3^{4/{3}^*} S_3^{4/3}(\tilde{R}_2^{2/{3}^*}\tilde{R}_2^{2/3}) \to \tau^+\tau^- b \bar b$ processes. 

We focus on the scenario presented in Sec.~\ref{subsection:fourY} when $S_3$ is at the TeV scale and $\tilde R_2$ is assumed not to feed significantly in the $\tau^+\tau^- b \bar b$ signal. In this case the CMS bound can be applied directly to the $S_3^{4/{3}}$ state, at a benchmark mass of $m_{S_3}=1$\,TeV, decaying into a $\tau b$ pair with a BR given by
\begin{equation}
\beta\approx\frac{|y_{b\tau}|^2}{|y_{b\tau}|^2+|y_{s\tau}|^2}\,.
\end{equation}
\noindent Here we neglect the small widths of $S_3^{4/3}$ into both muonic channels. Results are given in Fig.~\ref{fig:2LQrecast} (left panel), where the dashed blue contours represent the 95\% C.L.\ exclusion limits for different LHC luminosities and the red region represents the $1\,\sigma$ region for the low-energy fit derived in Sec.~\ref{subsection:fourY}. It is worth mentioning that we did not include other contributions which could potentially tighten these bounds, for example, contributions from $pp \to  S_3^{4/{3}^*} S_3^{4/3} \to  \tau^+\tau^- b s$.\footnote{This process will produce events in the signal region defined in Ref.~\cite{Sirunyan:2017yrk}, which is based on only one $b$-tagged jet and not two. We have also excluded from our analysis contributions coming from the non-QCD LQ pair production.} The search starts losing sensitivity for $m_{S_3}$ above 1\,TeV, while for masses larger than $1.2$\,TeV the search does not produce any useful limits. In conclusion, a third generation LQ pair production search at the LHC is not a sensitive probe for this particular flavor structure of our LQ model.

We now turn to the scenario where two generic third generation LQs (like e.g. $S_3$ and $\tilde R_2$) are at the TeV scale and both contribute to LQ pair production. A naive reinterpretation of the CMS limits~\cite{Sirunyan:2017yrk} can be performed when (i) both LQ components are degenerate in mass\footnote{For the non-degenerate case the results of Ref.~\cite{Sirunyan:2017yrk} are not directly applicable given that each LQ will have different kinematic distributions leading to different selection efficiencies in the signal region.}, (ii) interference terms between the final state $\tau b$ pairs at the amplitude level are negligible and, (iii) the LQ decay widths are small enough in order to guarantee the narrow width approximation (NWA) assumed in the experimental search. This scenario would thus correspond to the particular case we investigated in Sec.~\ref{subsection:sixY}. 

As shown in Appendix~\ref{Appendix_A}, the CMS bound for one LQ can be directly mapped into a bound for two degenerate LQs we denote with $\mathrm{LQ}_1$ and $\mathrm{LQ}_2$, for simplicity, when the associated BRs are $\beta_1$ and $\beta_2$, respectively. (See Fig.~2 (right panel) of Ref.~\cite{Sirunyan:2017yrk} for the experimental limit.) The inferred limits, which apply in general to two third-generation LQs with non-interfering final states, are presented in Fig.~\ref{fig:2LQrecast} (right panel) for an integrated luminosity of 12.9\,fb$^{-1}$ (solid green contour). For example, we find that both LQs with equal masses bellow 930\,GeV (600\,GeV) are excluded at 95\%\,C.L.\ if $\beta_{\text{eff}}=1$ ($\beta_{\text{eff}}=0.5$), where $\beta_{\text{eff}}=\sqrt{(\beta_1^2+\beta_2^2)/2}$. We also include LHC projections for an integrated luminosity of 300\,fb$^{-1}$ (dashed green contour) by rescaling the CMS Collaboration limits with the square root of the luminosity ratio. At this projected luminosity the LQ pair production search is not sensitive for masses above $1.3$\,TeV. These results, again, can be used for our LQ model for degenerate $S_3$ and $\tilde{R}_2$ LQs at the TeV scale. Unfortunately, the LHC bounds on the couplings extracted from LQ pair production search are not strong enough to probe this scenario either.

\subsection{High-mass $\tau\tau$ production}

In this section we study the implication of light $S_3$ and $\tilde{R}_2$ leptoquarks for high-$p_T$ $\tau\tau$ production at the LHC. It was shown in Refs.~\cite{Faroughy:2016osc,Greljo:2015mma} that $\tau\tau$ resonance searches at the LHC produce stringent constrains on a large class of models explaining the $R_{D^{(*)}}$ anomaly. In what follows we give predictions for the deviation from the SM in the invariant mass tails of $pp\to\tau^+\tau^-$ and derive bounds at different luminosities for the parameter space of the present LQ model from the 13\,TeV ATLAS Collaboration resonance search at 3.2\,fb$^{-1}$~\cite{Aaboud:2016cre}.

\begin{figure}[t!]
\centering
\includegraphics[width=0.9\textwidth]{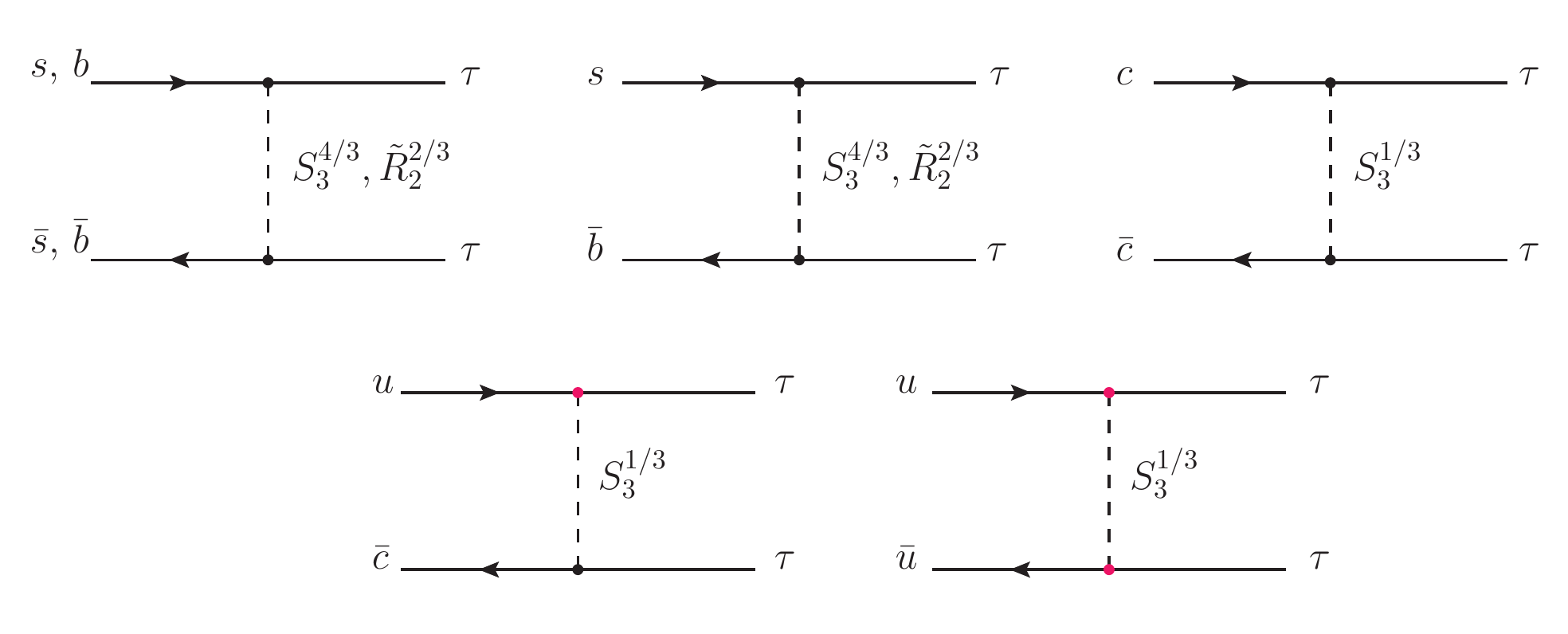}
\caption{\label{fig:feyndiag}Leading order Feynman diagrams for the $t$-channel $S_3$ and $\tilde{R}_2$ exchanges in $pp \to \tau^+\tau^-$ process. The red vertex indicates the presence of the $|V_{us}|$ Cabibbo suppression in the coupling.}
\end{figure}

Both LQs contribute to $pp\to\tau^+\tau^-$ production exclusively through Yukawa interactions by exchanging $S_3^{4/3}$, $S_3^{1/3}$, and $\tilde{R}_2^{2/3}$ components in the $t$-channel from partonic $q\bar q$ annihilation. The relevant Feynman diagrams are depicted in Fig.~\ref{fig:feyndiag}. Potentially large contributions may come from the processes with incoming strange quarks $s\bar s\to \tau^+\tau^-$ and $s\bar b\, (b \bar s) \to \tau^+\tau^-$, followed by sub-leading contributions from bottom, charm and up quark initiated processes $b\bar b\, (c\bar c)\,(u\bar u)\to\tau^+\tau^-$. The flavor structure in Eq.~\eqref{eq:textureS3} also allows for $S_3^{1/3}$ to couple to $u$ and $\tau$ via the CKM mixing. Nevertheless, this coupling is proportional to $|V_{us}|\,y_{s\tau}$, meaning that $\tau\tau$ production from incoming up quarks is Cabibbo suppressed leading to negligible cross-sections of order $|V_{us}|^2$ and $|V_{us}|^4$ for the processes $c\bar u \to \tau^+\tau^-$ and $u\bar u \to \tau^+\tau^-$, respectively. The Cabibbo suppressed vertices are shown in red in Feynman diagrams of Fig.~\ref{fig:feyndiag}. On the other hand, at high-$x$ the large proton PDF of the valence up quark in the process $u\bar c \to \tau^+\tau^-$ can marginally compensate for the $|V_{us}|$ suppression in the amplitude giving a contribution comparable to $c\bar c \to \tau^+\tau^-$ in the total cross-section.

 We now focus on the total cross-section $\sigma^{\text{fid}}_{\text{TOT}}$ of $pp\to\tau^+\tau^-$ far from the $Z$-pole in the high-mass tails of the $\tau\tau$ invariant mass distribution. We will, for definiteness, study the scenario where only $S_3$ contributes to $\tau\tau$ production. The couplings of $\tilde R_2$ are assumed to be small and can thus be safely neglected for this collider study. This is in accordance with the outcome of the numerical study presented in Sec.~\ref{subsection:fourY}. 
 
 At leading-order (LO), $\tau\tau$ production will receive contributions from the $t$-channel exchange of $S_3$, from the $s$-channel SM Drell-Yan $p p \to Z/\gamma^*\to\tau\tau$ production, and from interference effects between these processes. The high-mass kinematic region is defined by the following fiducial cuts on the final states: $p_T> 100$\,GeV (50\,GeV) for the leading (sub-leading) $\tau$ and a high invariant mass cut for the $\tau\tau$ pair of $m_{\tau\tau}>600$\,GeV. We define the {\it signal strength} $\mu_{pp\to\tau\tau}$ as the ratio of $\sigma^{\text{fid}}_{\text{TOT}}$ with the the SM Drell-Yan fiducial cross-section $\sigma^{\text{fid}}_{\text{SM}}$:

\begin{equation}\label{eq:signal-strength}
\mu_{pp\to\tau\tau}\ \equiv\ \sigma^{\text{fid}}_{\text{TOT}}\,/\,\sigma^{\text{fid}}_{\text{SM}}\ =\ 1\ +\ \sigma^{\text{fid}}_{\text{LQ}}\,/\,\sigma^{\text{fid}}_{\text{SM}}\,.
\end{equation}

\noindent Here the fiducial cross-section $\sigma^{\text{fid}}_{\text{LQ}}$ includes all NP contributions from both the LQ squared and LQ-SM interference amplitudes, i.e., $\sigma^{\text{fid}}_{\text{LQ}}=2\,\text{Re}(\mathcal{A}_{\text{SM}}^*\mathcal{A}_{\text{LQ}})\,+\,|\mathcal{A}_{\text{LQ}}|^2$. The ratio $\sigma^{\text{fid}}_{\text{LQ}}\,/\,\sigma^{\text{fid}}_{\text{SM}}$ quantifies the NP deviation of the total fiducial cross-section from the expected SM prediction. The LQ Yukawa couplings enter in $\sigma^{\text{fid}}_{\text{LQ}}$ as

\begin{equation}\label{fidxsec}
\sigma^{\text{fid}}_{\text{LQ}}(y_{s\tau},y_{b\tau})\ =\ \sigma_{s\bar s}\,(y_{s\tau})\ +\ \sigma_{s\bar b}\,(y_{s\tau},y_{b\tau})\ +\ \sigma_{b\bar b}\,(y_{b\tau})\ +\ \sigma_{c\bar c,u\bar u,u\bar c}\,(y_{s\tau}),
\end{equation}

\noindent In order to conform with the analysis in Sec.~\ref{sec:fit} we assume all Yukawa couplings to be real. Here $\sigma_{s\bar s}$, $\sigma_{s\bar b}$, $\sigma_{b\bar b}$ and  $\sigma_{c\bar c,u\bar u,u\bar c}$ correspond to the fiducial cross-sections of the processes $s\bar s\to\tau^+\tau^-$,  $s\bar b\, (\bar s b)\to\tau^+\tau^-$, $b\bar b\to\tau^+\tau^-$ and  $c\bar c\,(u\bar u)\,(u\bar c)\to\tau^+\tau^-$ respectively. These can be expressed as generic quartic polynomials in the couplings:
\begin{eqnarray}
\sigma_{s\bar s}\,(y_{s\tau})&=&y_{s\tau}^4\, A_1\ +\ y_{s\tau}^2\, B_1\,, \label{eq:xsec1}\\
\sigma_{s\bar b}\,(y_{s\tau},y_{b\tau}) &=& y_{s\tau}^2y_{b\tau}^2\, A_2\,, \label{eq:xsec2}\\
\sigma_{b\bar b}\,(y_{b\tau}) &=& y_{b\tau}^4\, A_3\ +\ y_{b\tau}^2\, B_3\,, \label{eq:xsec3}\\
\sigma_{c\bar c,u\bar u,u\bar c}\,(y_{s\tau})&=&y_{s\tau}^4\, A_4\ -\ y_{s\tau}^2\, B_4\, .\label{eq:xsec4}
\end{eqnarray}

\begin{figure}[b!]
\centering
\includegraphics[width=0.425\textwidth]{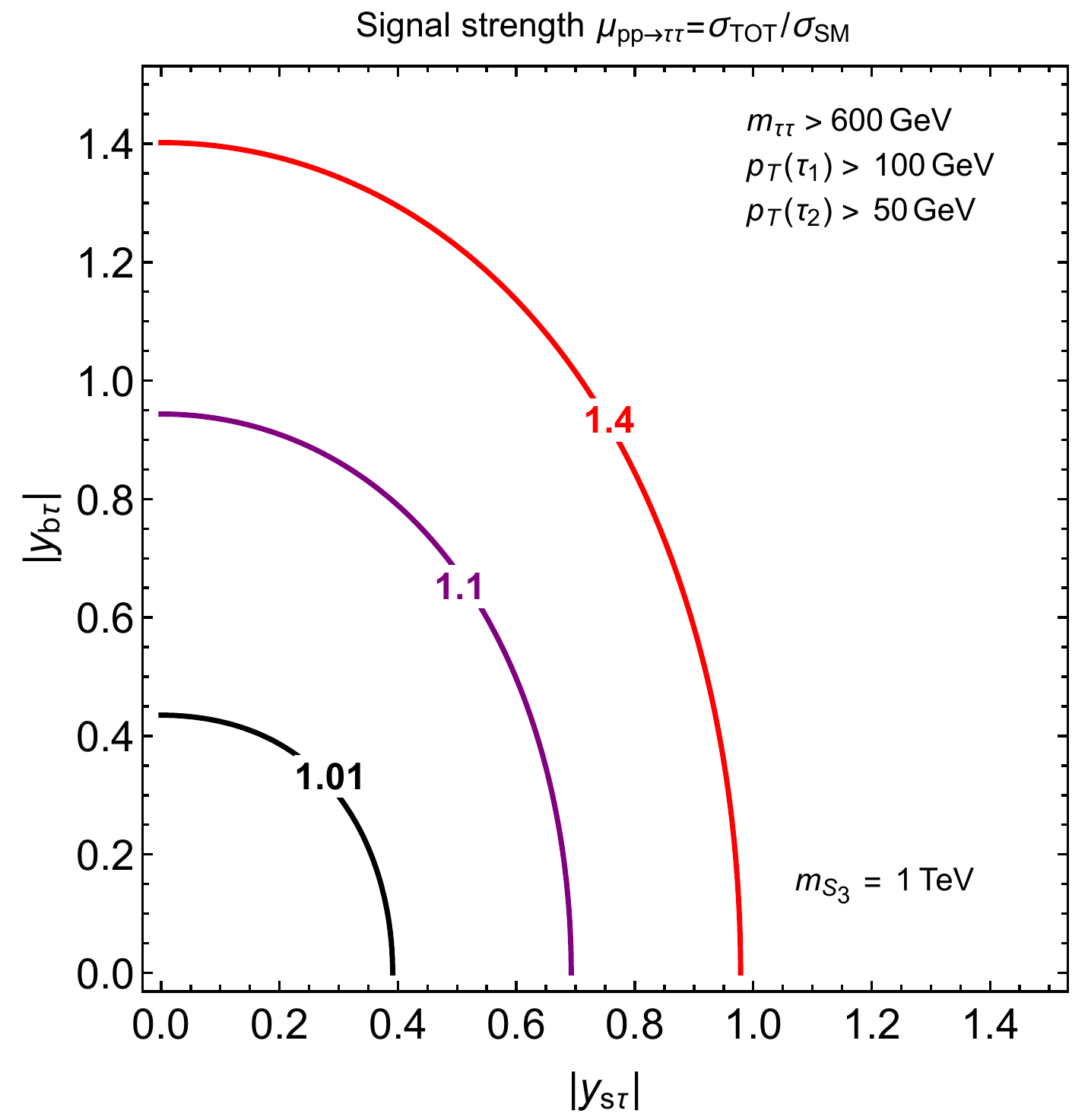}
\includegraphics[width=0.45\textwidth]{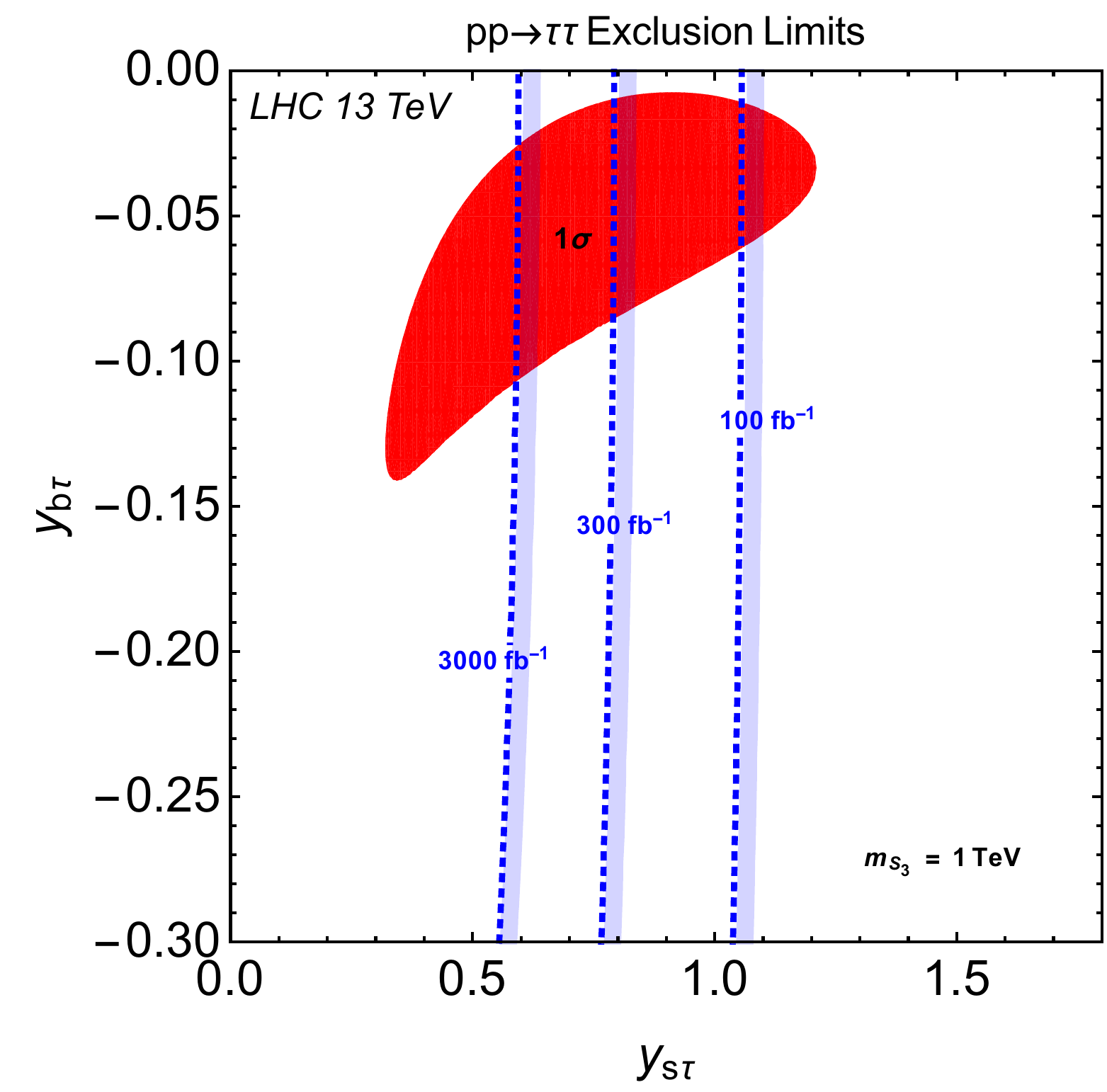}
\caption{\label{fig:fig1_a} (Left panel) Contours of constant signal strength $\mu_{pp\to\tau\tau}$ for different deviations from the SM prediction at $m_{S_3}=1$\,TeV. (Right panel) 95\% C.L.\ limits for LHC luminosities of 100, 300, and, 3000\,fb$^{-1}$ (dotted blue contours) from recasting high-mass $\tau\tau$ searches by ATLAS~\cite{Aaboud:2016cre} for $m_{S_3}=1$\,TeV. The red region corresponds to the 1\,$\sigma$ low-energy fit.}
\end{figure}

The polynomial coefficients $A_i$ and $B_i$ are functions of the mass $m_{S_3}$ describing the LQ squared amplitudes and LQ-SM interference amplitudes, respectively. In Eqs.~\eqref{eq:xsec1}--\eqref{eq:xsec4} we define all polynomial coefficients to be positive and include the explicit signs of the LQ-SM interference coefficients $B_i$, indicating the presence of either destructive or constructive interference amplitudes. The origin of the sign of these interference terms can be traced back to the SM amplitude proportional to the $T_3-\sin^2\theta_W\, Q$ coupling of the $Z$ boson with the incoming quarks along with the sign of the LQ Yukawa interactions of $S_3^{4/3\,(1/3)}$. The interference arising between the $Z$ boson and the $S_3^{4/3\,(1/3)}$ components is driven by the weak isospin of the quark doublets, i.e., positive (negative) for up(down)-type quarks. This translates into a constructive interference for down-type quarks in $d \bar d,\,s\bar s\,,b\bar b\to\tau\tau$ processes and destructive interference for up-type quarks in $c\bar c\,,u\bar u\to\tau\tau$ processes. As a consequence, this relative sign that appears between the LQ-SM interferences in different $\tau\tau$ production channels leads to a partial cancellation in the cross-section $\sigma^{\text{fid}}_{\text{LQ}}$. For more details see Appendix~\ref{Appendix_B}.
 
For the calculations below we choose a benchmark mass value of 1\,TeV. In order to extract the values of the polynomial coefficients $A_i$ and $B_i$ we generate in {\tt FeynRules}~\cite{Alloul:2013bka} the UFO model file for the Lagrangian for $S_3$ and simulate in {\tt MadGraph5}~\cite{Alwall:2014hca} 13\,TeV event samples of $pp\to\tau^+\tau^-$ subject to the high-mass fiducial cuts discussed above for different values of the couplings. The specific values of the coefficients can be found in Appendix~\ref{Appendix_B}. In Fig.~\ref{fig:fig1_a} (left panel) we show results for different contours of constant signal strength $\mu_{pp\to\tau\tau}$ in the $y_{s\tau}$--$y_{b\tau}$ plane for the 1\,TeV mass scenario. This shows that our LQ model equipped with the proposed flavor structure with $y_{s\tau,b\tau}$ couplings of $\mathcal{O}(1)$, predicts at the LHC an enhancement of $\mathcal{O}(10\%)$ in the $\tau\tau$ tails when compared to the SM.

In the remaining part of this section we confront our model with existing LHC data and extract $95\%$\,C.L.\ limits for the model parameters at different integrated luminosities. For this we recast a heavy $Z^\prime$ search by the ATLAS Collaboration with $3.2$\,fb$^{-1}$ of data in the fully hadronic $\tau_{\text{had}}\tau_{\text{had}}$ channel~\cite{Aaboud:2016cre}. For the recast we generate at LO in {\tt MadGraph5} a set of the LQ $\tau_{\text{had}}\tau_{\text{had}}$ signal samples followed by parton showering and hadronization in {\tt Pythia8}~\cite{Sjostrand:2014zea}. We have included interference effects between the LQ signals and the SM Drell-Yan background process. Detector effects are simulated in {\tt Delphes3}~\cite{deFavereau:2013fsa} with settings tuned according to the experimental environment of the $\tau_{\text{had}}\tau_{\text{had}}$ inclusive category as described in Ref.~\cite{Aaboud:2016cre}. For this category, events are selected if the reconstructed objects satisfies the following requirements: 
\begin{itemize}
\item $p_T> 110$ GeV\,(55\,GeV) for the leading (sub-leading) $\tau_{\text{had}}$.
\item Events with isolated electrons (muons) are vetoed if $p_T>10$\,GeV ($p_T>15$\,GeV).
\item Opposite sign $\tau_{\text{had}} \tau_{\text{had}}$ with back-to-back topology in the transverse plane, $\Delta\phi(\tau_{\text{had}} \tau_{\text{had}})>2.7$.
\item Total transverse mass cut of $m_T^{\text{tot}} > 350$\,GeV.
\end{itemize}
Here $m_T^{\text{tot}}$ is the dynamical variable used to reconstruct the invariant mass of the visible part of $\tau_{\text{had}}\tau_{\text{had}}$ defined as $({m_T^{\text{tot}}})^2\equiv m_T^2(\tau_1,\tau_2)+m_T^2(\tau_1,E^{\text{miss}}_T)+m_T^2(\tau_2,E^{\text{miss}}_T)$ where $E^{\text{miss}}_T$ is the total missing energy in the event and $m_T^2(A,B)=p_T(A)p_T(B)(1-\cos\Delta\phi(A,B))$ is the squared transverse mass of objects $A$ and $B$. In order to extract limits from the tails of the $m_T^{\text{tot}}$ distributions we use the statistical analysis presented in Ref.~\cite{Faroughy:2016osc}. For $m_{\mathrm{LQ}}>1$\,TeV the most sensitive bin corresponds to the last one ($m_T^{\text{tot}}>684$\,GeV) where a point in parameter space is excluded at 95\%\,C.L.\ if the number of events in that bin exceeds 11, 36 and 113 at integrated luminosities of 30, 300, 3000\,fb$^{-1}$ respectively. Here we have applied a naive scaling of the limits for the 3.2\,fb$^{-1}$ search to arbitrary luminosities with $\sqrt{\mathcal{L}_{\text{int}}/3.2\, \small{\text{fb}^{-1}}}$. The results are given in Fig.~\ref{fig:fig1_a} (Right panel) for the benchmark mass of $m_{\mathrm{LQ}}=1$\,TeV. There the regions to the right of the dotted blue boundaries are excluded at 95\%\,C.L.\ for future LHC luminosities of 30\,fb$^{-1}$,  300\,fb$^{-1}$ and 3000\,fb$^{-1}$. In the plot we have also included the $1\,\sigma$ region in solid red obtained from the low-energy fit to all flavor experiments performed in Sec.~\ref{subsection:fourY}. Notice that these bounds are conservative given that we have not included NLO QCD corrections and have not taken into account other processes such as non-resonant $qg\to \tau\tau q$ that can produce sizeable contributions to the inclusive $\tau_{\text{had}}\tau_{\text{had}}$ category \footnote{This last process, although $\alpha_s$ suppressed, has a large PDF from the initial gluon that enhances the total inclusive cross-section by $\mathcal{O}(10\%)$ or more.}. From these results we conclude that the High-Luminosity LHC can probe a large portion of the parameter space for the Yukawa coupling ansatz of Sec.~\ref{subsection:fourY}.

\section{GUT completion}
\label{sec:gut}

The preceding sections were devoted to the study of the impact of light scalar fields $S_3$ and $\tilde{R}_2$ with potentially sizeable couplings to the quark-lepton pairs on the flavor physics processes and LHC observables. Here we want to demonstrate that these fields and associated couplings can originate from a consistent grand unified theory (GUT) model.

To insure that the LHC accessible leptoquarks $S_3$ and $\tilde{R}_2$ are not in conflict with stringent limits on matter stability it is necessary that both $S_3$ and $\tilde{R}_2$ do not couple to the quark-quark pairs either directly or through the mixing with other scalars in a specific model of unification. It turns out that one can meet this requirement in an $SU(5)$ model that comprises $5$-, $15$-, $24$-, and $45$-dimensional scalar representations~\cite{Dorsner:2017wwn}. The decomposition of the scalar sector of that model is $\bm{5}= (\Psi_D,
\Psi_T) = (\bm{1},\bm{2},1/2)\oplus(\bm{3},\bm{1},-1/3),$ $
\bm{15}= (\Phi_a, \Phi_b, \Phi_c) = (\bm{1},\bm{3},1)\oplus
(\bm{3},\bm{2},1/6)\oplus(\bm{6},\bm{1},-2/3)$, $\bm{24}=(\Sigma_8
, \Sigma_3, \Sigma_{(3,2)}, \Sigma_{(\overline{3},2)},
\Sigma_{24}) = (\bm{8},\bm{1},0)\oplus(\bm{1},\bm{3},0)
\oplus(\bm{3},\bm{2},-5/6)\oplus(\overline{\bm{3}},\bm{2},5/6)
\oplus(\bm{1},\bm{1},0)$, and $\bm{45}=(\Delta_1, \Delta_2, \Delta_3,
\Delta_4, \Delta_5, \Delta_6, \Delta_7) =
(\bm{8},\bm{2},1/2)\oplus (\overline{\bm{6}},\bm{1}, -1/3) \oplus
(\bm{3},\bm{3},-1/3) \oplus (\overline{\bm{3}}, \bm{2}, -7/6)
\oplus (\bm{3},\bm{1}, -1/3) \oplus (\overline{\bm{3}}, \bm{1},
4/3) \oplus (\bm{1}, \bm{2}, 1/2)$, where $\Phi_b$ and $\Delta_3$ are identified with $\tilde{R}_2$ and $S^*_3$, respectively. The fermions of the SM, on the other hand, are embedded within the tenplets and fiveplets in the usual manner~\cite{Georgi:1974sy}.

We first show that this GUT scenario is compatible with the viable gauge coupling unification. To this end, we take all scalar fields in the model that mediate proton decay at tree-level to reside at or above $10^{12}$\,GeV. These fields are $\Psi_T$, $\Delta_5$, and $\Delta_6$~\cite{Dorsner:2017wwn}. We furthermore set the masses of both $S_3$ and $\tilde{R}_2$ at $1$\,TeV and constrain all remaining scalar fields to be at or above one scale we simply denote $m$ that is to be determined through the requirement that the gauge coupling unification takes place at the one-loop level. Note that $\Sigma_{24}$ does not affect unification. Also, $\Sigma_{(3,2)}$ and $\Sigma_{(\overline{3},2)}$ are not physical fields since they provide necessary degrees of freedom for the baryon and lepton number violating gauge bosons $X$ and $Y$ of the $SU(5)$ origin to become massive fields. 

The gauge couplings meet at the unification scale $m_{\mathrm{GUT}}$ when the following equation is satisfied~\cite{Giveon:1991zm}
\begin{equation}
\label{eq:a}
\frac{B_{23}}{B_{12}}=\frac{5}{8}
\frac{\sin^2
\theta_W-\alpha/\alpha_S}{3/8-\sin^2 \theta_W}=0.721 \pm 0.004,
\end{equation}
where the right-hand side is evaluated using $\alpha_S(m_Z)=0.1193\pm0.0016$, $\alpha^{-1}(m_Z)=127.906\pm0.019$, and
$\sin^2 \theta_W=0.23126\pm0.00005$~\cite{Agashe:2014kda}. The left-hand side depends on the particle content and the mass spectrum of the model. Namely, coefficients $B_{ij}$ are $B_{ij}=\sum_{J} (b^J_{i}-b^J_{j}) r_{J}$, where $b^J_{i}$ are the well-known $\beta$-function coefficients of particle $J$ with mass $m_J$ and $r_J=(\ln
m_{\mathrm{GUT}}/m_{J})/(\ln m_{\mathrm{GUT}}/m_{Z})$. The sum goes through all particles beside the SM ones that reside between $Z$ boson mass $m_{Z}$ and $m_\mathrm{GUT}$. The convention is such that $b^J_{1}$, $b^J_{2}$, and $b^J_{3}$ are associated with $U(1)$, $SU(2)$, and $SU(3)$ of the SM, respectively. We identify $m_\mathrm{GUT}$ not only with the gauge coupling unification scale but with the masses of the proton decay mediating gauge boson fields $X$ and $Y$.

If and when unification takes place for a given $m$ we evaluate $m_\mathrm{GUT}$ using equation~\cite{Giveon:1991zm} 
\begin{equation}
\label{eq:b}
\ln \frac{m_{\mathrm{GUT}}}{m_Z}=\frac{16 \pi}{5
\alpha} \frac{3/8-\sin^2 \theta_W}{B_{12}}=\frac{184.8 \pm
0.1}{B_{12}}
\end{equation}
to check that $m_\mathrm{GUT} \geq 5 \times 10^{15}$\,GeV in order to satisfy stringent bounds on the $X$ and $Y$ gauge boson mediated proton decay. To actually set a lower bound on $m$ we fix $m_\mathrm{GUT} = 5 \times 10^{15}$\,GeV in our analysis and maximise $m$. We find that $m = 3.1 \times 10^{10}$\,GeV when the masses of both $\tilde{R}_2$ and $S_3$ are at $1$\,TeV. The masses of all other scalar particles in the model are $m_{\Psi_D}=10^2$\,GeV,
$m_{\Psi_T}=10^{12}$\,GeV, $m_{\Phi_a}=m_\mathrm{GUT}$, $m_{\Phi_c}=m$, $m_{\Sigma_8}=m$
, $m_{\Sigma_3}=m_\mathrm{GUT}$, $m_{\Delta_1}=m$, $m_{\Delta_2}=m$, $m_{\Delta_4}=1.2 \times 10^{12}$\,GeV, $m_{\Delta_5}=10^{12}$\,GeV, and $m_{\Delta_6}=10^{12}$\,GeV, $m_{\Delta_7}= m_\mathrm{GUT}$. Note that the SM Higgs is in principle a mixture of $\Psi_T$ and $\Delta_7$. We accordingly take one state to be light and treat the mass of the other as a free parameter that is between $m$ and $m_\mathrm{GUT}$. 

The fact that viable unification can take place when $S_3$ and $\tilde{R}_2$ are both light does not come as a surprise. Note that the SM field content yields $B^{\textrm{SM}}_{23}/B^{\textrm{SM}}_{12}=0.53$ instead of the experimentally required value given in Eq.~\eqref{eq:a}. The nice feature of the set-up with light $S_3$ and $\tilde{R}_2$ is that both fields have positive $b^J_{23}$ and negative $b^J_{12}$ coefficients. This not only helps in bringing the left-hand side of Eq.~\eqref{eq:a} in agreement with the required experimental value but simultaneously raises the GUT scale $m_{\mathrm{GUT}}$ through Eq.~\eqref{eq:b}. The relevant coefficients are $b^{S_3}_{23}=9/6$, $b^{S_3}_{12}=-27/15$, $b^{\tilde{R}_2}_{23}=1/6$, and $b^{\tilde{R}_2}_{12}=-7/15$. Again, our findings demonstrate that the gauge coupling unification is possible for light $S_3$ and $\tilde{R}_2$ in this particular model.

We next demonstrate that the explicit forms of the Yukawa couplings of $S_3$ and $\tilde{R}_2$ that are used in Sec.~\ref{sec:fit} to produce numerical fits can originate from the appropriate $SU(5)$ operators. It is also argued that the model can accommodate realistic masses of the SM fermions.

The $S_3 \in \overline{\bm{45}}$ lepton-quark couplings originate from the $SU(5)$ contraction $y^{45}_{ij} \bm{10}_i \overline{\bm{5}}_j \overline{\bm{45}}$, where $\bm{10}_i$ are the usual fermionic tenplets, $i(=1,2,3)$ is the generation index, and $y^{45}$ is a $3 \times 3$ matrix in flavor space. We can thus identify $y$ of Eq.~\eqref{eq:S3} with $y^{45}/\sqrt{2}$, where $y^{45}$ is related to the difference in masses between charged fermions and down-type quarks~\cite{Dorsner:2017wwn}. This follows from the fact that there are actually two operators that contribute towards the charged fermion and down-type quark masses in this $SU(5)$ model~\cite{Georgi:1979df}. One is $y^{45}_{ij} \bm{10}_i \overline{\bm{5}}_j \overline{\bm{45}}$ and the other is $y^{5}_{ij} \bm{10}_i \overline{\bm{5}}_j \overline{\bm{5}}$, where $y^{5}$ is an arbitrary complex matrix. Clearly, $y^{45}=\sqrt{2} y$ and $y^{5}$ together contain enough parameters to easily address observed mismatch between the charged fermion and down-type quark masses. For completeness we specify that the up-type quark masses originate from a single contraction $x_{ij} \bm{10}_i \bm{10}_j \bm{5}$, where $x_{ij}$ is a symmetric complex $3 \times 3$ matrix. 

The $\tilde{R}_2 \in \bm{15}$ lepton-quark couplings are symmetric in flavor space since they originate from $y^{15}_{ij} \overline{\bm{5}}_i \overline{\bm{5}}_j \bm{15}$, where $\overline{\bm{5}}_i$ are the usual fermionic fiveplets. We identify $\tilde{y}_{ij}$ of Eq.~\eqref{eq:R2} with $-(D_R y^{15})_{ij }/\sqrt{2}$ in the physical basis for the down-type quarks and charged leptons, where $D_R$ represents unitary transformation of the right-chiral down-type quarks. If we take that $y^{15}_{33} \neq 0$ we obtain the form of $\tilde{y}$ that is used in the fit of Sec.~\ref{sec:fit} when we consider joint effect of $S_3$ and $\tilde{R}_2$ on flavor observables.

It is worth mentioning that it is possible to address neutrino masses within this model. Namely, if one turns on a vacuum expectation value of the electrically neutral field $\Phi_a \in \bm{15}$ one can generate neutrino masses of Majorana nature via type II see-saw mechanism~\cite{Lazarides:1980nt,Mohapatra:1980yp} through the same operator that yields the $\tilde{R}_2$ lepton-quark couplings, i.e., $y^{15}_{ij} \overline{\bm{5}}_i \overline{\bm{5}}_j \bm{15}$. In this particular instance the entries in $y^{15}$ would need to be responsible for the observed mass-squared differences and mixing angles in the neutrino sector. That requirement would not be compatible with a simple ansatz for the structure of $\tilde{y}$ given in Eq.~\eqref{eq:textureR2}. Again, viable neutrino masses would only be possible if we depart from that ansatz and assume that the $\tilde{y}_{ij}$ entries are sufficiently small to avoid flavor constraints for light $\tilde{R}_2$. This is in agreement with the findings we presented in Sec.~\ref{subsection:sixY}. Note that the neutrino Majorana masses could also receive partial contribution through the one-loop processes, where the particles in the loop are down-quarks and a mixture of $S_3$ and $\tilde{R}_2$~\cite{Chua:1999si,Mahanta:1999xd,Dorsner:2017wwn}.   

The preceding discussion demonstrates that the $SU(5)$ GUT model comprising $5$-, $15$-, $24$-, and $45$-dimensional scalar representations, with the canonical embedding of the SM fermions, can accommodate light $S_3$ and $\tilde{R}_2$ and describe observed fermion masses of the SM without any conflict with relevant experimental constraints.

\section{Conclusion}
\label{sec:conclusion}

Our aim, in the present work, is to
accommodate the observed lepton non-universality in charged current
processes, signalled by $R_{D^{(*)}}$, as well as lepton
non-universality and the global tension in the $b \to s \mu^+ \mu^-$
sector through the introduction of light scalar LQ $S_3$. This LQ
emerges naturally in the context of a specific $SU(5)$ GUT model and
has to be accompanied by another light scalar LQ $\tilde R_2$ which improves gauge coupling unification and aids neutrino mass generation. 

The first
state, $S_3$, couples left-handed $s$ and $b$ to left-handed 
$\mu$ and $\tau$ and is capable of accommodating $b\to s\ell^+ \ell^-$
sector. Because of its weak triplet nature it also couples to
up-type quarks and neutrinos which are precisely the additional couplings needed
to address $R_{D^{(*)}}$. Large couplings needed for $R_{D^{(*)}}$ cause the weak triplet $S_3$
to inevitably contribute to other well constrained flavor observables
that agree with the SM predictions. We have analyzed those in detail and
demonstrated that the most pressing ones, 
$R_{\nu\nu}^{(*)} = \br(B \to K ^{(*)} \bar \nu\nu)/\br(B \to K ^{(*)} \bar
\nu\nu)_\mrm{SM}$ and $\Delta m_s$, allow only for minor improvement
of $R_{D^{(*)}}$ puzzle. We furthermore show that the second state, $\tilde R_2$,
cannot significantly improve the agreement with data. 

 Based on the numerical values of the LQ Yukawa couplings as
 obtained in the flavor fit we recast two LQ collider
 searches: (i) search for pair produced LQs decaying to
 $b\tau b\tau$, (ii) search for high-mass $\tau\tau$ final state which
 is sensitive to the $t$-channel LQ exchange. From the recast
 of the search for the LQ pair production we 
 find that 
 the proposed scenario with $m_{S_3} = 1\e{TeV}$ 
 cannot be significantly probed in a large portion of the parameter
 space even with
 $300\e{fb}^{-1}$ of integrated luminosity at the Large Hadron
 Collider. Complementary searches for
 $\tau\tau$ final states produced via a single LQ exchange, on the
 other hand, are not as hampered by large LQ masses and are already
 excluding corners of the parameter space with largest individual
 Yukawa couplings. Moreover, since the flavor fit requires increasing
 Yukawa couplings for larger LQ masses the sensitivity of high-mass
 $\tau\tau$ final state search does not degrade at higher masses as
 compared to the pair production mechanism. This method can probe
 almost entire parameter space of the model at
 $3000\e{fb}^{-1}$ of integrated luminosity.

A natural ultraviolet completion for the two LQ states is an
$SU(5)$ GUT. We demonstrate that a particular
setting with $5$-, $15$-, $24$-, and $45$-dimensional scalar representations is consistent with unification of the gauge
couplings, where light $S_3$ and $\tilde R_2$ leptoquarks reside in $45$- and $15$-dimensional
representations, respectively. Furthermore, baryon number violation is sufficiently suppressed by lack
of diquark couplings of $S_3$ and high enough scale of the rest of the
GUT spectrum. The model also accommodates the masses of all fermions
of the SM.

\acknowledgments

We thank J.F.\ Kamenik and M.\ Nardecchia for insightful
discussions. N.K. would like to thank B.\ Capdevila and S.\
Descotes-Genon for kindly providing results of the global fit of $b\to s
\mu\mu$ in the scenario with two leptoquarks. This work has been
supported in part by Croatian Science Foundation under the project
7118. S.F.\ and N.K.\ acknowledge support of the Slovenian Research
Agency through research core funding No.\ P1-0035. I.D.\ acknowledges
support of COST Action CA15108. D.A.F. acknowledges support by the
'Young Researchers Programme' of the Slovenian Research
Agency.

\appendix

\section{LQ pair production recast}
\label{Appendix_A}
In this appendix we give a reinterpretation of the results by the CMS Collaboration~\cite{Sirunyan:2017yrk} for the case of two LQs, denoted LQ$_1$ and LQ$_2$. When addressing the LQ pair production from QCD interactions, a model with two LQs of degenerate mass $m_{\mathrm{LQ}} \equiv m_{\mathrm{LQ}_1} = m_{\mathrm{LQ}_2}$ with BRs $\beta_1$ and $\beta_2$ for LQ$_{1}\to\tau b$ and LQ$_{2}\to\tau b$, respectively, can be consistently mapped to a model with only one LQ, denoted here as $\mathrm{LQ}$, with an {\it effective} mass $m_{\text{eff}}$ and an {\it effective} BR $\beta_{\text{eff}}$ for LQ\,$\to\tau b$. In this case, assuming the NWA, the total cross-section for $pp \to \mathrm{LQ}^*\mathrm{LQ} \to \tau^+\tau^- b \bar b$ is factorized into production and decay modes as
\begin{equation}
\label{eq:1}
\sigma_{pp\to\tau\tau bb}=\beta_{\text{eff}}^2\times \sigma_{\text{pair}}(m_{\text{eff}}),
\end{equation}
where $\sigma_{\text{pair}}$ is the $pp\to \mathrm{LQ}^*\mathrm{LQ}$ pair production cross-section that depends exclusively on the LQ mass when only QCD interactions are taken into account. For the numerical calculations we use the approximate expression from Ref.~\cite{Mandal:2015lca} for the cross-section at NLO 
\begin{equation}\label{approxSig}
\sigma_{\text{pair}}(m)\approx\exp\Big\{\sum_{n=-2}^2\, C_n\,\Big(\frac{m}{\text{[TeV]}}\Big)^n \Big\} [\text{fb}]\,,
\end{equation}
where $(C_{-2},C_{-1},C_0,C_1,C_2)=(-0.300,3.318,2.762,-3.780,-0.299)$ at NLO in QCD for LHC collision energies of $\sqrt{s}=13$\,TeV. Equating the right hand side of Eq.~\eqref{eq:1} to the total cross-section derived in the two LQ scenario $\sigma_{pp\to\tau\tau bb} =(\beta_1^2+\beta_2^2)\, \sigma_{\text{pair}}(m_{\mathrm{LQ}})$ and demanding $0\le\beta_{\text{eff}}\le1$ we find 
\begin{equation}
\label{map2to1}
\beta_{\text{eff}}=\sqrt{\frac{\beta_1^2+\beta_2^2}{2}}\, , \ \ \ m_{\text{eff}}\,=\, \sigma^{-1}(\,2\,\sigma_{\text{pair}}(m_{\text{LQ}})\,)\, ,
\end{equation}
where $\sigma^{-1}$ is the inverse function of Eq.~\eqref{approxSig}. Here we assume negligible interference effects between the decay products of the LQ$_{1,2}$ and simply add two cross-sections together. After calculating $\sigma^{-1}$ numerically we can use Eq.~\eqref{map2to1} to map the CMS Collaboration 12.9\,fb$^{-1}$ exclusion limits in the $\beta$--$m_{\mathrm{LQ}}$ plane as reported in Fig.~9 of Ref.~\cite{Sirunyan:2017yrk} into the exclusion limits for two generic non-interfering third-generation LQs with degenerate mass. These limits are shown in Fig.~\ref{fig:2LQrecast}.

\section{High-mass $\tau\tau$ production cross-sections}
\label{Appendix_B}

We obtain the following fiducial cross-sections in fb for the process $pp \to \tau\tau$ for $m_{\mathrm{LQ}}=1$\,TeV:
\begin{eqnarray}
\sigma_{s\bar s}(y_{s\tau})&=&12.042 \, y_{st}^4 + 5.126\,y_{st}^2\,, \label{eq:Axsec1}\\
\sigma_{s\bar b}(y_{s\tau},y_{b\tau}) &=&12.568\, y_{s\tau}^2y_{b\tau}^2\,, \label{eq:Axsec2}\\
\sigma_{b\bar b}(y_{b\tau}) &=& 3.199\, y_{b\tau}^4 +1.385\,y_{b\tau}^2\,,\label{eq:Axsec3}\\
\sigma_{c\bar c,u\bar u,u\bar c}(y_{s\tau}) &=& 3.987 \,y_{s\tau}^4 -5.189\, y_{s\tau}^2\,.\label{eq:Axsec4}
\end{eqnarray}

 \noindent Notice that in each individual production channel the interferences can be large. In particular, these dominate in $c\bar c\,(u\bar u)(u\bar c)\to\tau\tau$ production over the squared LQ terms for Yukawa couplings of order one, as shown in Eq.~\eqref{eq:Axsec4}. Only after summing across all channels the total interference is found to be sub-leading when compared to the total LQ squared amplitudes in most portions of parameter space. This happens because of an accidental cancellation between the constructive $S_3$--$Z$ interference in $s\bar s\to \tau\tau$ given by the second term in Eq.~\eqref{eq:Axsec1} and the destructive $S_3$--$Z$ interference in $c\bar c\,(u\bar u)(u\bar c)\to \tau\tau$ given by the second term in Eq.~\eqref{eq:Axsec4}. The remaining small (constructive) interference after cancellations is mostly given by $\tau\tau$ production from bottom fusion and is negligible in high-mass $\tau\tau$ searches for the current level of experimental uncertainties.

\bibliography{refs}

\providecommand{\href}[2]{#2}\begingroup\raggedright\begin{thebibliography}{10}

\bibitem{Lees:2012xj}
{\scshape BaBar} collaboration, J.~P. Lees et~al., \emph{{Evidence for an
  excess of $\bar{B} \to D^{(*)} \tau^-\bar{\nu}_\tau$ decays}},
  \href{https://doi.org/10.1103/PhysRevLett.109.101802}{\emph{Phys. Rev. Lett.}
  {\bfseries 109} (2012) 101802},
  [\href{https://arxiv.org/abs/1205.5442}{{\ttfamily 1205.5442}}].

\bibitem{Lees:2013uzd}
{\scshape BaBar} collaboration, J.~P. Lees et~al., \emph{{Measurement of an
  Excess of $\bar{B} \to D^{(*)}\tau^- \bar{\nu}_\tau$ Decays and Implications
  for Charged Higgs Bosons}},
  \href{https://doi.org/10.1103/PhysRevD.88.072012}{\emph{Phys. Rev.}
  {\bfseries D88} (2013) 072012},
  [\href{https://arxiv.org/abs/1303.0571}{{\ttfamily 1303.0571}}].

\bibitem{Huschle:2015rga}
{\scshape Belle} collaboration, M.~Huschle et~al., \emph{{Measurement of the
  branching ratio of $\bar{B} \to D^{(\ast)} \tau^- \bar{\nu}_\tau$ relative to
  $\bar{B} \to D^{(\ast)} \ell^- \bar{\nu}_\ell$ decays with hadronic tagging
  at Belle}}, \href{https://doi.org/10.1103/PhysRevD.92.072014}{\emph{Phys.
  Rev.} {\bfseries D92} (2015) 072014},
  [\href{https://arxiv.org/abs/1507.03233}{{\ttfamily 1507.03233}}].

\bibitem{Adachi:2009qg}
{\scshape Belle} collaboration, I.~Adachi et~al., \emph{{Measurement of $B \to
  D^{(*)} \tau \nu$ using full reconstruction tags}},  in \emph{{Proceedings,
  24th International Symposium on Lepton-Photon Interactions at High Energy
  (LP09): Hamburg, Germany, August 17-22, 2009}}, 2009,
  \href{https://arxiv.org/abs/0910.4301}{{\ttfamily 0910.4301}},
  \href{https://inspirehep.net/record/834881/files/arXiv:0910.4301.pdf}{https://inspirehep.net/record/834881/files/arXiv:0910.4301.pdf}.

\bibitem{Bozek:2010xy}
{\scshape Belle} collaboration, A.~Bozek et~al., \emph{{Observation of $B^+ \to
  \bar D^{*0} \tau^+ \nu_\tau$ and Evidence for $B^+ \to \bar D^0 \tau^+
  \nu_\tau$ at Belle}},
  \href{https://doi.org/10.1103/PhysRevD.82.072005}{\emph{Phys. Rev.}
  {\bfseries D82} (2010) 072005},
  [\href{https://arxiv.org/abs/1005.2302}{{\ttfamily 1005.2302}}].

\bibitem{Aaij:2015yra}
{\scshape LHCb} collaboration, R.~Aaij et~al., \emph{{Measurement of the ratio
  of branching fractions $\mathcal{B}(\bar{B}^0 \to
  D^{*+}\tau^{-}\bar{\nu}_{\tau})/\mathcal{B}(\bar{B}^0 \to
  D^{*+}\mu^{-}\bar{\nu}_{\mu})$}},
  \href{https://doi.org/10.1103/PhysRevLett.115.159901,
  10.1103/PhysRevLett.115.111803}{\emph{Phys. Rev. Lett.} {\bfseries 115}
  (2015) 111803}, [\href{https://arxiv.org/abs/1506.08614}{{\ttfamily
  1506.08614}}].

\bibitem{Hirose:2016wfn}
{\scshape Belle} collaboration, S.~Hirose et~al., \emph{{Measurement of the
  $\tau$ lepton polarization and $R(D^*)$ in the decay $\bar{B} \to D^* \tau^-
  \bar{\nu}_\tau$}},
  \href{https://doi.org/10.1103/PhysRevLett.118.211801}{\emph{Phys. Rev. Lett.}
  {\bfseries 118} (2017) 211801},
  [\href{https://arxiv.org/abs/1612.00529}{{\ttfamily 1612.00529}}].

\bibitem{Becirevic:2016yqi}
D.~Bečirević, S.~Fajfer, N.~Košnik and O.~Sumensari, \emph{{Leptoquark model
  to explain the $B$-physics anomalies, $R_K$ and $R_D$}},
  \href{https://doi.org/10.1103/PhysRevD.94.115021}{\emph{Phys. Rev.}
  {\bfseries D94} (2016) 115021},
  [\href{https://arxiv.org/abs/1608.08501}{{\ttfamily 1608.08501}}].

\bibitem{Bigi:2016mdz}
D.~Bigi and P.~Gambino, \emph{{Revisiting $B\to D \ell \nu$}},
  \href{https://doi.org/10.1103/PhysRevD.94.094008}{\emph{Phys. Rev.}
  {\bfseries D94} (2016) 094008},
  [\href{https://arxiv.org/abs/1606.08030}{{\ttfamily 1606.08030}}].

\bibitem{Amhis:2016xyh}
Y.~Amhis et~al., \emph{{Averages of $b$-hadron, $c$-hadron, and $\tau$-lepton
  properties as of summer 2016}},
  \href{https://arxiv.org/abs/1612.07233}{{\ttfamily 1612.07233}}.

\bibitem{Fajfer:2012jt}
S.~Fajfer, J.~F. Kamenik, I.~Nisandzic and J.~Zupan, \emph{{Implications of
  Lepton Flavor Universality Violations in B Decays}},
  \href{https://doi.org/10.1103/PhysRevLett.109.161801}{\emph{Phys. Rev. Lett.}
  {\bfseries 109} (2012) 161801},
  [\href{https://arxiv.org/abs/1206.1872}{{\ttfamily 1206.1872}}].

\bibitem{Ligeti:2016npd}
Z.~Ligeti, M.~Papucci and D.~J. Robinson, \emph{{New Physics in the Visible
  Final States of $B\to D^{(*)}\tau\nu$}},
  \href{https://doi.org/10.1007/JHEP01(2017)083}{\emph{JHEP} {\bfseries 01}
  (2017) 083}, [\href{https://arxiv.org/abs/1610.02045}{{\ttfamily
  1610.02045}}].

\bibitem{Crivellin:2016ejn}
A.~Crivellin, J.~Fuentes-Martin, A.~Greljo and G.~Isidori, \emph{{Lepton Flavor
  Non-Universality in B decays from Dynamical Yukawas}},
  \href{https://doi.org/10.1016/j.physletb.2016.12.057}{\emph{Phys. Lett.}
  {\bfseries B766} (2017) 77--85},
  [\href{https://arxiv.org/abs/1611.02703}{{\ttfamily 1611.02703}}].

\bibitem{Altmannshofer:2017poe}
W.~Altmannshofer, P.~S.~B. Dev and A.~Soni, \emph{{$R_{D^{(*)}}$ anomaly: A
  possible hint for natural supersymmetry with $R$-parity violation}},
  \href{https://arxiv.org/abs/1704.06659}{{\ttfamily 1704.06659}}.

\bibitem{Alonso:2016oyd}
R.~Alonso, B.~Grinstein and J.~Martin~Camalich, \emph{{Lifetime of $B_c^-$
  Constrains Explanations for Anomalies in $B\to D^{(*)}\tau\nu$}},
  \href{https://doi.org/10.1103/PhysRevLett.118.081802}{\emph{Phys. Rev. Lett.}
  {\bfseries 118} (2017) 081802},
  [\href{https://arxiv.org/abs/1611.06676}{{\ttfamily 1611.06676}}].

\bibitem{Crivellin:2014zpa}
A.~Crivellin and S.~Pokorski, \emph{{Can the differences in the determinations
  of $V_{ub}$ and $V_{cb}$ be explained by New Physics?}},
  \href{https://doi.org/10.1103/PhysRevLett.114.011802}{\emph{Phys. Rev. Lett.}
  {\bfseries 114} (2015) 011802},
  [\href{https://arxiv.org/abs/1407.1320}{{\ttfamily 1407.1320}}].

\bibitem{Bhattacharya:2014wla}
B.~Bhattacharya, A.~Datta, D.~London and S.~Shivashankara, \emph{{Simultaneous
  Explanation of the $R_K$ and $R(D^{(*)})$ Puzzles}},
  \href{https://doi.org/10.1016/j.physletb.2015.02.011}{\emph{Phys. Lett.}
  {\bfseries B742} (2015) 370--374},
  [\href{https://arxiv.org/abs/1412.7164}{{\ttfamily 1412.7164}}].

\bibitem{Bhattacharya:2015ida}
S.~Bhattacharya, S.~Nandi and S.~K. Patra, \emph{{Optimal-observable analysis
  of possible new physics in $B\to D^{(\ast)}\tau\nu_{\tau}$}},
  \href{https://doi.org/10.1103/PhysRevD.93.034011}{\emph{Phys. Rev.}
  {\bfseries D93} (2016) 034011},
  [\href{https://arxiv.org/abs/1509.07259}{{\ttfamily 1509.07259}}].

\bibitem{Hati:2015awg}
C.~Hati, G.~Kumar and N.~Mahajan, \emph{{$\bar{B}\rightarrow D^{(\ast)}\tau
  \bar{\nu}$ excesses in ALRSM constrained from $B$, $D$ decays and
  $D^{0}-\bar{D}^{0}$ mixing}},
  \href{https://doi.org/10.1007/JHEP01(2016)117}{\emph{JHEP} {\bfseries 01}
  (2016) 117}, [\href{https://arxiv.org/abs/1511.03290}{{\ttfamily
  1511.03290}}].

\bibitem{Sakaki:2014sea}
Y.~Sakaki, M.~Tanaka, A.~Tayduganov and R.~Watanabe, \emph{{Probing New Physics
  with $q^2$ distributions in $\bar{B} \to D^{(*)} \tau \bar\nu$}},
  \href{https://doi.org/10.1103/PhysRevD.91.114028}{\emph{Phys. Rev.}
  {\bfseries D91} (2015) 114028},
  [\href{https://arxiv.org/abs/1412.3761}{{\ttfamily 1412.3761}}].

\bibitem{Capdevila:2017ert}
B.~Capdevila, S.~Descotes-Genon, L.~Hofer and J.~Matias, \emph{{Hadronic
  uncertainties in $B \to K^* \mu^+ \mu^-$: a state-of-the-art analysis}},
  \href{https://doi.org/10.1007/JHEP04(2017)016}{\emph{JHEP} {\bfseries 04}
  (2017) 016}, [\href{https://arxiv.org/abs/1701.08672}{{\ttfamily
  1701.08672}}].

\bibitem{Aaij:2014ora}
{\scshape LHCb} collaboration, R.~Aaij et~al., \emph{{Test of lepton
  universality using $B^{+}\rightarrow K^{+}\ell^{+}\ell^{-}$ decays}},
  \href{https://doi.org/10.1103/PhysRevLett.113.151601}{\emph{Phys. Rev. Lett.}
  {\bfseries 113} (2014) 151601},
  [\href{https://arxiv.org/abs/1406.6482}{{\ttfamily 1406.6482}}].

\bibitem{Aaij:2017vbb}
{\scshape LHCb} collaboration, R.~Aaij et~al., \emph{{Test of lepton
  universality with $B^{0} \rightarrow K^{*0}\ell^{+}\ell^{-}$ decays}},
  \href{https://arxiv.org/abs/1705.05802}{{\ttfamily 1705.05802}}.

\bibitem{Bordone:2016gaq}
M.~Bordone, G.~Isidori and A.~Pattori, \emph{{On the Standard Model predictions
  for $R_K$ and $R_{K^*}$}},
  \href{https://doi.org/10.1140/epjc/s10052-016-4274-7}{\emph{Eur. Phys. J.}
  {\bfseries C76} (2016) 440},
  [\href{https://arxiv.org/abs/1605.07633}{{\ttfamily 1605.07633}}].

\bibitem{Hiller:2003js}
G.~Hiller and F.~Kruger, \emph{{More model-independent analysis of $b \to s$
  processes}}, \href{https://doi.org/10.1103/PhysRevD.69.074020}{\emph{Phys.
  Rev.} {\bfseries D69} (2004) 074020},
  [\href{https://arxiv.org/abs/hep-ph/0310219}{{\ttfamily hep-ph/0310219}}].

\bibitem{Wehle:2016yoi}
{\scshape Belle} collaboration, S.~Wehle et~al., \emph{{Lepton-Flavor-Dependent
  Angular Analysis of $B\to K^\ast \ell^+\ell^-$}},
  \href{https://doi.org/10.1103/PhysRevLett.118.111801}{\emph{Phys. Rev. Lett.}
  {\bfseries 118} (2017) 111801},
  [\href{https://arxiv.org/abs/1612.05014}{{\ttfamily 1612.05014}}].

\bibitem{Altmannshofer:2014cfa}
W.~Altmannshofer, S.~Gori, M.~Pospelov and I.~Yavin, \emph{{Quark flavor
  transitions in $L_\mu-L_\tau$ models}},
  \href{https://doi.org/10.1103/PhysRevD.89.095033}{\emph{Phys. Rev.}
  {\bfseries D89} (2014) 095033},
  [\href{https://arxiv.org/abs/1403.1269}{{\ttfamily 1403.1269}}].

\bibitem{Datta:2013kja}
A.~Datta, M.~Duraisamy and D.~Ghosh, \emph{{Explaining the $B \to K^\ast \mu^+
  \mu^-$ data with scalar interactions}},
  \href{https://doi.org/10.1103/PhysRevD.89.071501}{\emph{Phys. Rev.}
  {\bfseries D89} (2014) 071501},
  [\href{https://arxiv.org/abs/1310.1937}{{\ttfamily 1310.1937}}].

\bibitem{Hiller:2014yaa}
G.~Hiller and M.~Schmaltz, \emph{{$R_K$ and future $b \to s \ell \ell$ physics
  beyond the standard model opportunities}},
  \href{https://doi.org/10.1103/PhysRevD.90.054014}{\emph{Phys. Rev.}
  {\bfseries D90} (2014) 054014},
  [\href{https://arxiv.org/abs/1408.1627}{{\ttfamily 1408.1627}}].

\bibitem{Glashow:2014iga}
S.~L. Glashow, D.~Guadagnoli and K.~Lane, \emph{{Lepton Flavor Violation in $B$
  Decays?}}, \href{https://doi.org/10.1103/PhysRevLett.114.091801}{\emph{Phys.
  Rev. Lett.} {\bfseries 114} (2015) 091801},
  [\href{https://arxiv.org/abs/1411.0565}{{\ttfamily 1411.0565}}].

\bibitem{Gripaios:2014tna}
B.~Gripaios, M.~Nardecchia and S.~A. Renner, \emph{{Composite leptoquarks and
  anomalies in $B$-meson decays}},
  \href{https://doi.org/10.1007/JHEP05(2015)006}{\emph{JHEP} {\bfseries 05}
  (2015) 006}, [\href{https://arxiv.org/abs/1412.1791}{{\ttfamily 1412.1791}}].

\bibitem{Greljo:2015mma}
A.~Greljo, G.~Isidori and D.~Marzocca, \emph{{On the breaking of Lepton Flavor
  Universality in B decays}},
  \href{https://doi.org/10.1007/JHEP07(2015)142}{\emph{JHEP} {\bfseries 07}
  (2015) 142}, [\href{https://arxiv.org/abs/1506.01705}{{\ttfamily
  1506.01705}}].

\bibitem{Ghosh:2014awa}
D.~Ghosh, M.~Nardecchia and S.~A. Renner, \emph{{Hint of Lepton Flavour
  Non-Universality in $B$ Meson Decays}},
  \href{https://doi.org/10.1007/JHEP12(2014)131}{\emph{JHEP} {\bfseries 12}
  (2014) 131}, [\href{https://arxiv.org/abs/1408.4097}{{\ttfamily 1408.4097}}].

\bibitem{Crivellin:2015mga}
A.~Crivellin, G.~D'Ambrosio and J.~Heeck, \emph{{Explaining
  $h\to\mu^\pm\tau^\mp$, $B\to K^* \mu^+\mu^-$ and $B\to K \mu^+\mu^-/B\to K
  e^+e^-$ in a two-Higgs-doublet model with gauged $L_\mu-L_\tau$}},
  \href{https://doi.org/10.1103/PhysRevLett.114.151801}{\emph{Phys. Rev. Lett.}
  {\bfseries 114} (2015) 151801},
  [\href{https://arxiv.org/abs/1501.00993}{{\ttfamily 1501.00993}}].

\bibitem{Crivellin:2015lwa}
A.~Crivellin, G.~D'Ambrosio and J.~Heeck, \emph{{Addressing the LHC flavor
  anomalies with horizontal gauge symmetries}},
  \href{https://doi.org/10.1103/PhysRevD.91.075006}{\emph{Phys. Rev.}
  {\bfseries D91} (2015) 075006},
  [\href{https://arxiv.org/abs/1503.03477}{{\ttfamily 1503.03477}}].

\bibitem{Crivellin:2017zlb}
A.~Crivellin, D.~Müller and T.~Ota, \emph{{Simultaneous Explanation of
  $R(D^{(*)})$ and $b\to s\mu^+\mu^-$: The Last Scalar Leptoquarks Standing}},
  \href{https://arxiv.org/abs/1703.09226}{{\ttfamily 1703.09226}}.

\bibitem{Sierra:2015fma}
D.~Aristizabal~Sierra, F.~Staub and A.~Vicente, \emph{{Shedding light on the
  $b\to s$ anomalies with a dark sector}},
  \href{https://doi.org/10.1103/PhysRevD.92.015001}{\emph{Phys. Rev.}
  {\bfseries D92} (2015) 015001},
  [\href{https://arxiv.org/abs/1503.06077}{{\ttfamily 1503.06077}}].

\bibitem{Varzielas:2015iva}
I.~de~Medeiros~Varzielas and G.~Hiller, \emph{{Clues for flavor from rare
  lepton and quark decays}},
  \href{https://doi.org/10.1007/JHEP06(2015)072}{\emph{JHEP} {\bfseries 06}
  (2015) 072}, [\href{https://arxiv.org/abs/1503.01084}{{\ttfamily
  1503.01084}}].

\bibitem{Crivellin:2015era}
A.~Crivellin, L.~Hofer, J.~Matias, U.~Nierste, S.~Pokorski and J.~Rosiek,
  \emph{{Lepton-flavour violating $B$ decays in generic $Z'$ models}},
  \href{https://doi.org/10.1103/PhysRevD.92.054013}{\emph{Phys. Rev.}
  {\bfseries D92} (2015) 054013},
  [\href{https://arxiv.org/abs/1504.07928}{{\ttfamily 1504.07928}}].

\bibitem{Celis:2015ara}
A.~Celis, J.~Fuentes-Martin, M.~Jung and H.~Serodio, \emph{{Family nonuniversal
  $Z'$ models with protected flavor-changing interactions}},
  \href{https://doi.org/10.1103/PhysRevD.92.015007}{\emph{Phys. Rev.}
  {\bfseries D92} (2015) 015007},
  [\href{https://arxiv.org/abs/1505.03079}{{\ttfamily 1505.03079}}].

\bibitem{Freytsis:2015qca}
M.~Freytsis, Z.~Ligeti and J.~T. Ruderman, \emph{{Flavor models for $\bar{B}
  \to D^{(*)} \tau \bar{\nu}$}},
  \href{https://doi.org/10.1103/PhysRevD.92.054018}{\emph{Phys. Rev.}
  {\bfseries D92} (2015) 054018},
  [\href{https://arxiv.org/abs/1506.08896}{{\ttfamily 1506.08896}}].

\bibitem{Fajfer:2015ycq}
S.~Fajfer and N.~Košnik, \emph{{Vector leptoquark resolution of $R_K$ and
  $R_{D^{(*)}}$ puzzles}},
  \href{https://doi.org/10.1016/j.physletb.2016.02.018}{\emph{Phys. Lett.}
  {\bfseries B755} (2016) 270--274},
  [\href{https://arxiv.org/abs/1511.06024}{{\ttfamily 1511.06024}}].

\bibitem{Cox:2016epl}
P.~Cox, A.~Kusenko, O.~Sumensari and T.~T. Yanagida, \emph{{SU(5) Unification
  with TeV-scale Leptoquarks}},
  \href{https://doi.org/10.1007/JHEP03(2017)035}{\emph{JHEP} {\bfseries 03}
  (2017) 035}, [\href{https://arxiv.org/abs/1612.03923}{{\ttfamily
  1612.03923}}].

\bibitem{Becirevic:2017jtw}
D.~Bečirević and O.~Sumensari, \emph{{A leptoquark model to accommodate
  $R_K^\mathrm{exp} <R_K^\mathrm{SM} $ and $R_{K^\ast}^\mathrm{exp}<
  R_{K^\ast}^\mathrm{SM}$}},
  \href{https://arxiv.org/abs/1704.05835}{{\ttfamily 1704.05835}}.

\bibitem{Kamenik:2017tnu}
J.~F. Kamenik, Y.~Soreq and J.~Zupan, \emph{{Lepton flavor universality
  violation without new sources of quark flavor violation}},
  \href{https://arxiv.org/abs/1704.06005}{{\ttfamily 1704.06005}}.

\bibitem{Arnan:2017lxi}
P.~Arnan, D.~Bečirević, F.~Mescia and O.~Sumensari, \emph{{Two Higgs Doublet
  Models and $b\to s$ exclusive decays}},
  \href{https://arxiv.org/abs/1703.03426}{{\ttfamily 1703.03426}}.

\bibitem{Ghosh:2017ber}
D.~Ghosh, \emph{{Explaining the $R_K$ and $R_{K^*}$ anomalies}},
  \href{https://arxiv.org/abs/1704.06240}{{\ttfamily 1704.06240}}.

\bibitem{Bardhan:2016uhr}
D.~Bardhan, P.~Byakti and D.~Ghosh, \emph{{A closer look at the R$_{D}$ and
  R$_{D^*}$ anomalies}},
  \href{https://doi.org/10.1007/JHEP01(2017)125}{\emph{JHEP} {\bfseries 01}
  (2017) 125}, [\href{https://arxiv.org/abs/1610.03038}{{\ttfamily
  1610.03038}}].

\bibitem{Capdevila:2017bsm}
B.~Capdevila, A.~Crivellin, S.~Descotes-Genon, J.~Matias and J.~Virto,
  \emph{{Patterns of New Physics in $b\to s\ell^+\ell^-$ transitions in the
  light of recent data}},  \href{https://arxiv.org/abs/1704.05340}{{\ttfamily
  1704.05340}}.

\bibitem{Calibbi:2015kma}
L.~Calibbi, A.~Crivellin and T.~Ota, \emph{{Effective Field Theory Approach to
  $b\to s \ell \ell^{(')}$, $B\to K^{(*)} \nu \bar{\nu}$ and $B\to D^{(*)}\tau
  \nu$ with Third Generation Couplings}},
  \href{https://doi.org/10.1103/PhysRevLett.115.181801}{\emph{Phys. Rev. Lett.}
  {\bfseries 115} (2015) 181801},
  [\href{https://arxiv.org/abs/1506.02661}{{\ttfamily 1506.02661}}].

\bibitem{Dorsner:2016wpm}
I.~Doršner, S.~Fajfer, A.~Greljo, J.~F. Kamenik and N.~Košnik, \emph{{Physics
  of leptoquarks in precision experiments and at particle colliders}},
  \href{https://doi.org/10.1016/j.physrep.2016.06.001}{\emph{Phys. Rept.}
  {\bfseries 641} (2016) 1--68},
  [\href{https://arxiv.org/abs/1603.04993}{{\ttfamily 1603.04993}}].

\bibitem{Dorsner:2017wwn}
I.~Doršner, S.~Fajfer and N.~Košnik, \emph{{Leptoquark mechanism of neutrino
  masses within the grand unification framework}},
  \href{https://doi.org/10.1140/epjc/s10052-017-4987-2}{\emph{Eur. Phys. J.}
  {\bfseries C77} (2017) 417},
  [\href{https://arxiv.org/abs/1701.08322}{{\ttfamily 1701.08322}}].

\bibitem{Alonso:2015sja}
R.~Alonso, B.~Grinstein and J.~Martin~Camalich, \emph{{Lepton universality
  violation and lepton flavor conservation in $B$-meson decays}},
  \href{https://doi.org/10.1007/JHEP10(2015)184}{\emph{JHEP} {\bfseries 10}
  (2015) 184}, [\href{https://arxiv.org/abs/1505.05164}{{\ttfamily
  1505.05164}}].

\bibitem{Feruglio:2017rjo}
F.~Feruglio, P.~Paradisi and A.~Pattori, \emph{{On the Importance of
  Electroweak Corrections for B Anomalies}},
  \href{https://arxiv.org/abs/1705.00929}{{\ttfamily 1705.00929}}.

\bibitem{Feruglio:2016gvd}
F.~Feruglio, P.~Paradisi and A.~Pattori, \emph{{Revisiting Lepton Flavor
  Universality in B Decays}},
  \href{https://doi.org/10.1103/PhysRevLett.118.011801}{\emph{Phys. Rev. Lett.}
  {\bfseries 118} (2017) 011801},
  [\href{https://arxiv.org/abs/1606.00524}{{\ttfamily 1606.00524}}].

\bibitem{Becirevic:2015asa}
D.~Bečirević, S.~Fajfer and N.~Košnik, \emph{{Lepton flavor nonuniversality
  in $b\to s \ell^+ \ell^-$ processes}},
  \href{https://doi.org/10.1103/PhysRevD.92.014016}{\emph{Phys. Rev.}
  {\bfseries D92} (2015) 014016},
  [\href{https://arxiv.org/abs/1503.09024}{{\ttfamily 1503.09024}}].

\bibitem{Descotes-Genon:2015uva}
S.~Descotes-Genon, L.~Hofer, J.~Matias and J.~Virto, \emph{{Global analysis of
  $b\to s\ell\ell$ anomalies}},
  \href{https://doi.org/10.1007/JHEP06(2016)092}{\emph{JHEP} {\bfseries 06}
  (2016) 092}, [\href{https://arxiv.org/abs/1510.04239}{{\ttfamily
  1510.04239}}].

\bibitem{Abdesselam:2017kjf}
{\scshape Belle} collaboration, A.~Abdesselam et~al., \emph{{Precise
  determination of the CKM matrix element $\left| V_{cb}\right|$ with $\bar B^0
  \to D^{*\,+} \, \ell^- \, \bar \nu_\ell$ decays with hadronic tagging at
  Belle}},  \href{https://arxiv.org/abs/1702.01521}{{\ttfamily 1702.01521}}.

\bibitem{Glattauer:2015teq}
{\scshape Belle} collaboration, R.~Glattauer et~al., \emph{{Measurement of the
  decay $B\to D\ell\nu_\ell$ in fully reconstructed events and determination of
  the Cabibbo-Kobayashi-Maskawa matrix element $|V_{cb}|$}},
  \href{https://doi.org/10.1103/PhysRevD.93.032006}{\emph{Phys. Rev.}
  {\bfseries D93} (2016) 032006},
  [\href{https://arxiv.org/abs/1510.03657}{{\ttfamily 1510.03657}}].

\bibitem{Olive:2016xmw}
{\scshape Particle Data Group} collaboration, C.~Patrignani et~al.,
  \emph{{Review of Particle Physics}},
  \href{https://doi.org/10.1088/1674-1137/40/10/100001}{\emph{Chin. Phys.}
  {\bfseries C40} (2016) 100001}.

\bibitem{Cirigliano:2007xi}
V.~Cirigliano and I.~Rosell, \emph{{Two-loop effective theory analysis of $\pi
  (K) \to e \bar \nu_e [\gamma]$ branching ratios}},
  \href{https://doi.org/10.1103/PhysRevLett.99.231801}{\emph{Phys. Rev. Lett.}
  {\bfseries 99} (2007) 231801},
  [\href{https://arxiv.org/abs/0707.3439}{{\ttfamily 0707.3439}}].

\bibitem{Pich:2013lsa}
A.~Pich, \emph{{Precision Tau Physics}},
  \href{https://doi.org/10.1016/j.ppnp.2013.11.002}{\emph{Prog. Part. Nucl.
  Phys.} {\bfseries 75} (2014) 41--85},
  [\href{https://arxiv.org/abs/1310.7922}{{\ttfamily 1310.7922}}].

\bibitem{Decker:1994ea}
R.~Decker and M.~Finkemeier, \emph{{Short and long distance effects in the
  decay $\tau \to \pi \nu_\tau (\gamma)$}},
  \href{https://doi.org/10.1016/0550-3213(95)00597-L}{\emph{Nucl. Phys.}
  {\bfseries B438} (1995) 17--53},
  [\href{https://arxiv.org/abs/hep-ph/9403385}{{\ttfamily hep-ph/9403385}}].

\bibitem{Aaltonen:2014hua}
{\scshape CDF} collaboration, T.~A. Aaltonen et~al., \emph{{Study of Top-Quark
  Production and Decays involving a Tau Lepton at CDF and Limits on a
  Charged-Higgs Boson Contribution}},
  \href{https://doi.org/10.1103/PhysRevD.89.091101}{\emph{Phys. Rev.}
  {\bfseries D89} (2014) 091101},
  [\href{https://arxiv.org/abs/1402.6728}{{\ttfamily 1402.6728}}].

\bibitem{Aubert:2009ag}
{\scshape BaBar} collaboration, B.~Aubert et~al., \emph{{Searches for Lepton
  Flavor Violation in the Decays $\tau^\pm \to e^\pm \gamma$ and $\tau^\pm \to
  \mu^\pm \gamma$}},
  \href{https://doi.org/10.1103/PhysRevLett.104.021802}{\emph{Phys. Rev. Lett.}
  {\bfseries 104} (2010) 021802},
  [\href{https://arxiv.org/abs/0908.2381}{{\ttfamily 0908.2381}}].

\bibitem{Dorsner:2013tla}
I.~Doršner, S.~Fajfer, N.~Košnik and I.~Nišandžić, \emph{{Minimally
  flavored colored scalar in $\bar B \to D^{(*)} \tau \bar \nu$ and the mass
  matrices constraints}},
  \href{https://doi.org/10.1007/JHEP11(2013)084}{\emph{JHEP} {\bfseries 11}
  (2013) 084}, [\href{https://arxiv.org/abs/1306.6493}{{\ttfamily 1306.6493}}].

\bibitem{Queiroz:2014pra}
F.~S. Queiroz, K.~Sinha and A.~Strumia, \emph{{Leptoquarks, Dark Matter, and
  Anomalous LHC Events}},
  \href{https://doi.org/10.1103/PhysRevD.91.035006}{\emph{Phys. Rev.}
  {\bfseries D91} (2015) 035006},
  [\href{https://arxiv.org/abs/1409.6301}{{\ttfamily 1409.6301}}].

\bibitem{Lees:2012zz}
{\scshape BaBar} collaboration, J.~P. Lees et~al., \emph{{A search for the
  decay modes $B^{+-} \to h^{+-} \tau^{+-}l$}},
  \href{https://doi.org/10.1103/PhysRevD.86.012004}{\emph{Phys. Rev.}
  {\bfseries D86} (2012) 012004},
  [\href{https://arxiv.org/abs/1204.2852}{{\ttfamily 1204.2852}}].

\bibitem{Bailey:2015dka}
J.~A. Bailey et~al., \emph{{$B\to Kl^+l^-$ decay form factors from three-flavor
  lattice QCD}}, \href{https://doi.org/10.1103/PhysRevD.93.025026}{\emph{Phys.
  Rev.} {\bfseries D93} (2016) 025026},
  [\href{https://arxiv.org/abs/1509.06235}{{\ttfamily 1509.06235}}].

\bibitem{Gabbiani:1996hi}
F.~Gabbiani, E.~Gabrielli, A.~Masiero and L.~Silvestrini, \emph{{A Complete
  analysis of FCNC and CP constraints in general SUSY extensions of the
  standard model}},
  \href{https://doi.org/10.1016/0550-3213(96)00390-2}{\emph{Nucl. Phys.}
  {\bfseries B477} (1996) 321--352},
  [\href{https://arxiv.org/abs/hep-ph/9604387}{{\ttfamily hep-ph/9604387}}].

\bibitem{Buras:1990fn}
A.~J. Buras, M.~Jamin and P.~H. Weisz, \emph{{Leading and Next-to-leading {QCD}
  Corrections to $\epsilon$ Parameter and $B^0 - \bar{B}^0$ Mixing in the
  Presence of a Heavy Top Quark}},
  \href{https://doi.org/10.1016/0550-3213(90)90373-L}{\emph{Nucl. Phys.}
  {\bfseries B347} (1990) 491--536}.

\bibitem{Bazavov:2016nty}
{\scshape Fermilab Lattice, MILC} collaboration, A.~Bazavov et~al.,
  \emph{{$B^0_{(s)}$-mixing matrix elements from lattice QCD for the Standard
  Model and beyond}},
  \href{https://doi.org/10.1103/PhysRevD.93.113016}{\emph{Phys. Rev.}
  {\bfseries D93} (2016) 113016},
  [\href{https://arxiv.org/abs/1602.03560}{{\ttfamily 1602.03560}}].

\bibitem{Aoki:2016frl}
S.~Aoki et~al., \emph{{Review of lattice results concerning low-energy particle
  physics}}, \href{https://doi.org/10.1140/epjc/s10052-016-4509-7}{\emph{Eur.
  Phys. J.} {\bfseries C77} (2017) 112},
  [\href{https://arxiv.org/abs/1607.00299}{{\ttfamily 1607.00299}}].

\bibitem{Fajfer:2006av}
S.~Fajfer, J.~F. Kamenik and N.~Kosnik, \emph{{$b \to dd \bar s$ transition and
  constraints on new physics in $B^-$ decays}},
  \href{https://doi.org/10.1103/PhysRevD.74.034027}{\emph{Phys. Rev.}
  {\bfseries D74} (2006) 034027},
  [\href{https://arxiv.org/abs/hep-ph/0605260}{{\ttfamily hep-ph/0605260}}].

\bibitem{Altmannshofer:2009ma}
W.~Altmannshofer, A.~J. Buras, D.~M. Straub and M.~Wick, \emph{{New strategies
  for New Physics search in $B \to K^{*} \nu \bar{\nu}$, $B \to K \nu
  \bar{\nu}$ and $B \to X_{s} \nu \bar{\nu}$ decays}},
  \href{https://doi.org/10.1088/1126-6708/2009/04/022}{\emph{JHEP} {\bfseries
  04} (2009) 022}, [\href{https://arxiv.org/abs/0902.0160}{{\ttfamily
  0902.0160}}].

\bibitem{Buras:2014fpa}
A.~J. Buras, J.~Girrbach-Noe, C.~Niehoff and D.~M. Straub, \emph{{$ B\to
  {K}^{\left(\ast \right)}\nu \overline{\nu} $ decays in the Standard Model and
  beyond}}, \href{https://doi.org/10.1007/JHEP02(2015)184}{\emph{JHEP}
  {\bfseries 02} (2015) 184},
  [\href{https://arxiv.org/abs/1409.4557}{{\ttfamily 1409.4557}}].

\bibitem{Grygier:2017tzo}
{\scshape Belle} collaboration, J.~Grygier et~al., \emph{{Search for
  $\boldsymbol{B\to h\nu\bar{\nu}}$ decays with semileptonic tagging at
  Belle}},  \href{https://arxiv.org/abs/1702.03224}{{\ttfamily 1702.03224}}.

\bibitem{Greljo:2017vvb}
A.~Greljo and D.~Marzocca, \emph{{High-$p_T$ dilepton tails and flavor
  physics}}, \href{https://doi.org/10.1140/epjc/s10052-017-5119-8}{\emph{Eur.
  Phys. J.} {\bfseries C77} (2017) 548},
  [\href{https://arxiv.org/abs/1704.09015}{{\ttfamily 1704.09015}}].

\bibitem{deBoer:2015boa}
S.~de~Boer and G.~Hiller, \emph{{Flavor and new physics opportunities with rare
  charm decays into leptons}},
  \href{https://doi.org/10.1103/PhysRevD.93.074001}{\emph{Phys. Rev.}
  {\bfseries D93} (2016) 074001},
  [\href{https://arxiv.org/abs/1510.00311}{{\ttfamily 1510.00311}}].

\bibitem{Fajfer:2015mia}
S.~Fajfer and N.~Košnik, \emph{{Prospects of discovering new physics in rare
  charm decays}},
  \href{https://doi.org/10.1140/epjc/s10052-015-3801-2}{\emph{Eur. Phys. J.}
  {\bfseries C75} (2015) 567},
  [\href{https://arxiv.org/abs/1510.00965}{{\ttfamily 1510.00965}}].

\bibitem{Sirunyan:2017yrk}
{\scshape CMS} collaboration, A.~M. Sirunyan et~al., \emph{{Search for the
  third-generation scalar leptoquarks and heavy right-handed neutrinos in final
  states with two tau leptons and two jets in proton-proton collisions at
  $\sqrt{s}$ = 13 TeV}},  \href{https://arxiv.org/abs/1703.03995}{{\ttfamily
  1703.03995}}.

\bibitem{Faroughy:2016osc}
D.~A. Faroughy, A.~Greljo and J.~F. Kamenik, \emph{{Confronting lepton flavor
  universality violation in B decays with high-$p_T$ tau lepton searches at
  LHC}}, \href{https://doi.org/10.1016/j.physletb.2016.11.011}{\emph{Phys.
  Lett.} {\bfseries B764} (2017) 126--134},
  [\href{https://arxiv.org/abs/1609.07138}{{\ttfamily 1609.07138}}].

\bibitem{Aaboud:2016cre}
{\scshape ATLAS} collaboration, M.~Aaboud et~al., \emph{{Search for Minimal
  Supersymmetric Standard Model Higgs bosons $H/A$ and for a $Z^{\prime}$ boson
  in the $\tau \tau$ final state produced in $pp$ collisions at $\sqrt{s}=13$
  TeV with the ATLAS Detector}},
  \href{https://doi.org/10.1140/epjc/s10052-016-4400-6}{\emph{Eur. Phys. J.}
  {\bfseries C76} (2016) 585},
  [\href{https://arxiv.org/abs/1608.00890}{{\ttfamily 1608.00890}}].

\bibitem{Alloul:2013bka}
A.~Alloul, N.~D. Christensen, C.~Degrande, C.~Duhr and B.~Fuks,
  \emph{{FeynRules 2.0 - A complete toolbox for tree-level phenomenology}},
  \href{https://doi.org/10.1016/j.cpc.2014.04.012}{\emph{Comput. Phys. Commun.}
  {\bfseries 185} (2014) 2250--2300},
  [\href{https://arxiv.org/abs/1310.1921}{{\ttfamily 1310.1921}}].

\bibitem{Alwall:2014hca}
J.~Alwall, R.~Frederix, S.~Frixione, V.~Hirschi, F.~Maltoni, O.~Mattelaer
  et~al., \emph{{The automated computation of tree-level and next-to-leading
  order differential cross sections, and their matching to parton shower
  simulations}}, \href{https://doi.org/10.1007/JHEP07(2014)079}{\emph{JHEP}
  {\bfseries 07} (2014) 079},
  [\href{https://arxiv.org/abs/1405.0301}{{\ttfamily 1405.0301}}].

\bibitem{Sjostrand:2014zea}
T.~Sjöstrand, S.~Ask, J.~R. Christiansen, R.~Corke, N.~Desai, P.~Ilten et~al.,
  \emph{{An Introduction to PYTHIA 8.2}},
  \href{https://doi.org/10.1016/j.cpc.2015.01.024}{\emph{Comput. Phys. Commun.}
  {\bfseries 191} (2015) 159--177},
  [\href{https://arxiv.org/abs/1410.3012}{{\ttfamily 1410.3012}}].

\bibitem{deFavereau:2013fsa}
{\scshape DELPHES 3} collaboration, J.~de~Favereau, C.~Delaere, P.~Demin,
  A.~Giammanco, V.~Lemaître, A.~Mertens et~al., \emph{{DELPHES 3, A modular
  framework for fast simulation of a generic collider experiment}},
  \href{https://doi.org/10.1007/JHEP02(2014)057}{\emph{JHEP} {\bfseries 02}
  (2014) 057}, [\href{https://arxiv.org/abs/1307.6346}{{\ttfamily 1307.6346}}].

\bibitem{Georgi:1974sy}
H.~Georgi and S.~L. Glashow, \emph{{Unity of All Elementary Particle Forces}},
  \href{https://doi.org/10.1103/PhysRevLett.32.438}{\emph{Phys. Rev. Lett.}
  {\bfseries 32} (1974) 438--441}.

\bibitem{Giveon:1991zm}
A.~Giveon, L.~J. Hall and U.~Sarid, \emph{{SU(5) unification revisited}},
  \href{https://doi.org/10.1016/0370-2693(91)91289-8}{\emph{Phys. Lett.}
  {\bfseries B271} (1991) 138--144}.

\bibitem{Agashe:2014kda}
{\scshape Particle Data Group} collaboration, K.~A. Olive et~al., \emph{{Review
  of Particle Physics}},
  \href{https://doi.org/10.1088/1674-1137/38/9/090001}{\emph{Chin. Phys.}
  {\bfseries C38} (2014) 090001}.

\bibitem{Georgi:1979df}
H.~Georgi and C.~Jarlskog, \emph{{A New Lepton - Quark Mass Relation in a
  Unified Theory}},
  \href{https://doi.org/10.1016/0370-2693(79)90842-6}{\emph{Phys. Lett.}
  {\bfseries 86B} (1979) 297--300}.

\bibitem{Lazarides:1980nt}
G.~Lazarides, Q.~Shafi and C.~Wetterich, \emph{{Proton Lifetime and Fermion
  Masses in an SO(10) Model}},
  \href{https://doi.org/10.1016/0550-3213(81)90354-0}{\emph{Nucl. Phys.}
  {\bfseries B181} (1981) 287--300}.

\bibitem{Mohapatra:1980yp}
R.~N. Mohapatra and G.~Senjanovic, \emph{{Neutrino Masses and Mixings in Gauge
  Models with Spontaneous Parity Violation}},
  \href{https://doi.org/10.1103/PhysRevD.23.165}{\emph{Phys. Rev.} {\bfseries
  D23} (1981) 165}.

\bibitem{Chua:1999si}
C.-K. Chua, X.-G. He and W.-Y.~P. Hwang, \emph{{Neutrino mass induced
  radiatively by supersymmetric leptoquarks}},
  \href{https://doi.org/10.1016/S0370-2693(00)00325-7}{\emph{Phys. Lett.}
  {\bfseries B479} (2000) 224--229},
  [\href{https://arxiv.org/abs/hep-ph/9905340}{{\ttfamily hep-ph/9905340}}].

\bibitem{Mahanta:1999xd}
U.~Mahanta, \emph{{Neutrino masses and mixing angles from leptoquark
  interactions}}, \href{https://doi.org/10.1103/PhysRevD.62.073009}{\emph{Phys.
  Rev.} {\bfseries D62} (2000) 073009},
  [\href{https://arxiv.org/abs/hep-ph/9909518}{{\ttfamily hep-ph/9909518}}].

\bibitem{Mandal:2015lca}
T.~Mandal, S.~Mitra and S.~Seth, \emph{{Pair Production of Scalar Leptoquarks
  at the LHC to NLO Parton Shower Accuracy}},
  \href{https://doi.org/10.1103/PhysRevD.93.035018}{\emph{Phys. Rev.}
  {\bfseries D93} (2016) 035018},
  [\href{https://arxiv.org/abs/1506.07369}{{\ttfamily 1506.07369}}].

\end{thebibliography}\endgroup
\end{document}